\documentclass[11pt]{article}
\usepackage[margin=1in]{geometry}
\usepackage{amsmath,amssymb,amsthm,mathtools,array,longtable,booktabs}
\usepackage{fancyvrb}
\usepackage[T1]{fontenc}
\usepackage{lmodern}
\usepackage[hidelinks]{hyperref}
\usepackage{xurl}
\numberwithin{equation}{section}
\newtheorem{theorem}{Theorem}[section]
\newtheorem{lemma}[theorem]{Lemma}
\newtheorem{proposition}[theorem]{Proposition}
\newtheorem{corollary}[theorem]{Corollary}

\newcommand{\HH}{\mathcal H}
\newcommand{\BB}{\mathcal B}
\newcommand{\XX}{\mathcal X}
\newcommand{\dd}{\mathrm d}
\newcommand{\norm}[1]{\left\lVert #1\right\rVert}
\newcommand{\ip}[2]{\left\langle #1,#2\right\rangle}
\title{A computer-assisted proof of one-cell snaking and lattice trapping\\
for a nonlinear square lattice}

\author{Jiangchao Qian}
\author{
    Jiangchao Qian$^{1}$ \\[6pt]
    \small $^1$School of Engineering Science, Simon Fraser University,\\
    \small Burnaby, British Columbia V5A 4Y7, Canada \\[4pt]
    \small \texttt{jiangchao\_qian@sfu.ca}
}

\begin{document}
\maketitle
\begin{abstract}
For the interaction potential $\phi(t)=(1-e^{-3t^2})/6$, we construct
a connected branch of equilibria in the infinite square lattice with
crack far field $K\sqrt r\sin(\theta/2)$. A single lattice-cell segment
has exactly two nondegenerate quadratic folds and the full-space
stability sequence stable, index one, stable. The two fold loads obey
$0.45897<K_-<0.45923$ and $0.46219<K_+<0.46242$; in particular their
difference exceeds $0.00296$. A common-class uniqueness argument proves
exact translation closure and smooth continuation by lattice cells.
The proof combines energy-space arguments, analytic infinite-tail
estimates, interval inequalities, and exact integer positivity
certificates. All finite input data, verification programs, and saved
certificate outputs are supplied in the accompanying certificate archive (Appendix~\ref{app:archive}).
\end{abstract}
\tableofcontents

\section{Statement and interpretation}

Lattice trapping occurs when the discrete structure of a crystal permits a
crack to remain locally stable over a range of applied loads
\cite{thomson1971}. Buze, Hudson and Ortner studied finite-cell
approximations of atomistic crack equilibria and computed a sequence of
folds for the square-lattice potential used here \cite{buze2020}.
Their convergence analysis assumes a branch with specified regularity
and stability properties. Theorem~\ref{thm:main} addresses these
properties directly for one lattice cell and identifies its endpoints
under translation. For the numerical continuation of crack equilibria
in both antiplane and opening-mode geometries, see \cite{buze2021}.

Each undirected nearest-neighbor bond is counted once. The lattice is
$\Lambda=(\mathbb Z+1/2)^2=\mathbb Z^2-(1/2,1/2)$ and the potential is
\[
 \phi(t)=\frac{1-e^{-3t^2}}6.
\]
The reference norm omits the vertical negative-axis slit bonds, while
the physical energy and equations retain every nearest-neighbor
interaction. The precise energy space $\HH$, removal of constants,
convergent energy difference $E$, and force $F=E_v$ are defined below.
Write $h=\sqrt r\sin(\theta/2)$ with the cut on the negative axis, and
let $(\mathcal Tu)(x)=u(x-e_1)$.

\begin{theorem}[One-cell branch, folds, and translation closure]
\label{thm:main}
There are $T>0$ and a smooth regular branch
$s\mapsto(v(s),K(s))\in\HH\times(0,\infty)$, $s\in\mathbb R$,
such that $u(s)=K(s)h+v(s)$ is an exact equilibrium of the infinite
physical lattice, and
\[
 K(s+T)=K(s),\qquad u(s+T,x)=u(s,x-e_1)
\]
in the odd displacement gauge. Each half-open period has exactly two
simple quadratic folds. Starting at a stable point, the full Hessian
is coercive up to the first fold, has exactly one negative direction
between the folds, and is coercive after the second fold.

At each fold the Hessian is self-adjoint Fredholm of index zero, its
kernel is one-dimensional, and it is coercive on the orthogonal
complement of the kernel. For a unit kernel vector $q$, both
\[
 F_K[q]\ne0,\qquad F_{vv}[q,q][q]\ne0
\]
hold. The first fold is a local maximum $K_+$ of load, the second a
local minimum $K_-$, and
\[
 \frac{45897}{100000}<K_-<\frac{45923}{100000},\qquad
 \frac{46219}{100000}<K_+<\frac{46242}{100000},\qquad
 K_+-K_->\frac{37}{12500}>0.
\]
The one-cell segment is locally unique in the full energy space.
Its initial opening is $1/4$ and its terminal opening belongs to
$[0.4720462,0.4767331]$. Its two fold openings belong, respectively,
to $(29/100,8/25)$ and $(2/5,11/25)$.
\end{theorem}

The statement concerns this branch and a neighborhood of it; it does
not classify all equilibria. Stability is measured in the energy
space after quotienting constants, not by eigenvalues of a finite
Dirichlet matrix. The positive interval $(K_-,K_+)$ is the trapping
interval between these loss-of-stability thresholds. Translation
periodicity of configurations and load makes no claim of periodic
or zero energy release.

The proof has three parts. First we establish the analytic validation
and continuation implications. Next we give the finite certificates
and their infinite-tail estimates. Finally we combine those already
verified strict inequalities. The exact finite arrays and programs
are mathematical certificate data, supplied in the accompanying archive
described in Appendix~\ref{app:archive}; numerical equilibrium
solvers are not used as proof steps. Displayed decimal enclosures
are outward rational bounds. Numbers stored in binary arrays are
interpreted as exact dyadic rationals, with the normalization repairs
specified below.

\section{The lattice, the reference space, and the physical force}

Let
\[
 \Lambda=(\mathbb Z+\tfrac12)^2,
 \qquad \BB=\{(x,x+e_i):x\in\Lambda,\ i=1,2\},
 \qquad D_bu=u(y)-u(x)\quad(b=(x,y)).
\]
Thus every undirected nearest-neighbor bond occurs once. Set
\[
 c_n=\bigl((-n-\tfrac12,-\tfrac12),(-n-\tfrac12,\tfrac12)\bigr),
 \quad n\geq0,\qquad
 \Gamma=\{c_n:n\geq0\},\qquad \BB_0=\BB\setminus\Gamma.
\]
Only the reference norm omits $\Gamma$. The physical energy and force
always use the full set $\BB$. The interaction law is
\[
 \phi(t)=\frac{1-e^{-3t^2}}6,
 \qquad \tau(t)=\phi'(t)=te^{-3t^2},
 \qquad \phi''(t)=(1-6t^2)e^{-3t^2}.
\]
Write $p_+=(1/2,1/2)$, $p_-=(1/2,-1/2)$, and let $p=(p_-,p_+)$.
Define $\HH$ to be the completion of compactly supported functions
plus constants, modulo constants, in
\[
 \norm{v}_{\HH}^2=\sum_{b\in\BB_0}|D_bv|^2.
\]
The reference graph is connected. A representative is determined by
prescribing its value at one vertex. Differences of values at any two
fixed vertices are continuous linear functionals: a retained path of
length $m$ gives $|v(x)-v(y)|\leq\sqrt m\norm v_{\HH}$.
Consequently the completion consists of actual lattice functions modulo
constants, with the indicated differences.

\begin{lemma}[No finite-energy directions are excluded]
The space $\HH$ is exactly the set of lattice functions satisfying
$\sum_{b\in\BB_0}|D_bv|^2<\infty$, modulo constants.
\end{lemma}
\begin{proof}
The preceding path bounds identify every element of the completion
with a lattice function, and its gradient is its limiting
$\ell^2(\BB_0)$ gradient. Conversely, let $v$ be any lattice function
with finite retained energy. Choose a representative and truncate its
amplitude by
$T_M(t)=\max(-M,\min(t,M))$. For every bond,
$|D_b(T_M(v)-v)|\leq2|D_bv|$, and this difference tends to zero
as $M\to\infty$. Dominated convergence therefore gives
$T_M(v)\to v$ in the energy seminorm.

It remains to approximate a bounded finite-energy function $w$.
For $R>2$, use a radial cutoff $\vartheta_R$ equal to one on
$r\leq R$, equal to zero on $r\geq R^2$, and equal to
$\log(R^2/r)/\log R$ between these radii. Its bond differences
satisfy $|D_b\vartheta_R|\leq C/(r_b\log R)$ on the annulus,
with an adjusted constant on its two boundary layers. Counting
the $O(r)$ square-lattice bonds in each unit radial shell yields
\[
 \sum_{b\in\BB_0}|D_b\vartheta_R|^2
 \leq \frac{C}{\log R}.
\]
The omitted slit bonds only decrease this sum. The discrete product
rule, with either endpoint convention, now gives
\[
 \norm{\vartheta_Rw-w}_{\HH}^2
 \leq 2\sum_{b:\,r_b\geq R-1}|D_bw|^2
      +2\norm w_\infty^2\sum_b|D_b\vartheta_R|^2\longrightarrow0.
\]
Each $\vartheta_Rw$ has finite support. Applying this to $T_M(v)$
and then taking a diagonal sequence proves the reverse inclusion.
\end{proof}

Put
\[
 \ell(v)=\tfrac12D_pv,\qquad \XX=\ker\ell.
\]
Since $p\in\BB_0$, $\norm\ell\leq1/2$. Reflection in the horizontal
axis acts isometrically on $\HH$. The odd subspace $\HH_{\mathrm o}$
has a unique representative satisfying
$v(x_1,-x_2)=-v(x_1,x_2)$. On this subspace $\ell(v)=v(p_+)$.
We write $\XX_{\mathrm o}=\XX\cap\HH_{\mathrm o}$. Let $\zeta$ be
one at $p_+$, minus one at $p_-$, and zero elsewhere. Then
\[
 \ell(\zeta)=1,\qquad \norm\zeta_{\HH}^2=10.
\]
There is no simultaneous requirement that an arbitrary odd
representative vanish at $p_+$. When a full-space vector belongs to
$\XX$, however, its gauge can be chosen so that both pin values vanish.

\begin{lemma}[Crack traces]
For $v\in\HH$,
\[
 |D_{c_n}v|\leq\sqrt{2n+3}\norm v_{\HH},
 \qquad D_{c_n}v=o(\sqrt{n+1})\quad(n\to\infty).
\]
For $v\in\XX$ the first constant can be replaced by $\sqrt{2n+2}$.
In particular, for odd $v\in\XX$,
\[
 |v(-\tfrac12,\tfrac12)|\leq\frac{\norm v_{\HH}}{\sqrt2}.
\]
\end{lemma}
\begin{proof}
Join the endpoints of $c_n$ horizontally to the two pin vertices and
then use $p$. This retained path has $2n+3$ edges; Cauchy--Schwarz gives
the first bound. If $D_pv=0$, the pin contribution vanishes and the
same telescoping sum uses only the $2n+2$ horizontal edges. To obtain
the little-oh assertion, approximate $v$ in $\HH$ by a compactly
supported representative $v_j$. For sufficiently large $n$,
$D_{c_n}v_j=0$, and hence
$|D_{c_n}v|/\sqrt{n+1}\leq\sqrt3\norm{v-v_j}_{\HH}$.
First let $n$ tend to infinity and then $j$ tend to infinity. The last
claim uses $n=0$ and oddness.
\end{proof}

For $K>0$ prescribe
\[
 h(x)=\sqrt r\sin(\theta/2),\qquad U_K=Kh,
 \qquad r=|x|,\quad -\pi<\theta<\pi.
\]
The cut is the negative horizontal axis. The physical affine space is
$Kh+\HH$. Its load $K$ is uniquely determined, since $h\notin\HH$.
Indeed, along the upper negative-axis row, the squares of the
horizontal differences of $h$ are asymptotic to $1/(4n)$, so even this
subseries of its retained energy diverges. Distances in $\HH$ between
physical configurations will only be used after their loads have
been made equal.

\subsection{A divergence-free reference flux}

At integer dual vertices define $g(z)=\operatorname{Re}\sqrt z$, using
the principal square root and $g(0)=0$. For a positively oriented edge
with midpoint $z$, put
\[
 P_b=\begin{cases}
 g(z-e_2/2)-g(z+e_2/2),&b\parallel e_1,\\
 g(z+e_1/2)-g(z-e_1/2),&b\parallel e_2.
 \end{cases}
\]
The outgoing minus incoming fluxes cancel algebraically at every
lattice vertex. Moreover $P_{c_n}=0$. Therefore
\begin{equation}
 \sum_{b\in\BB}P_bD_bw=0
 \quad\hbox{for every compactly supported }w.
 \label{eq:core-1}
\end{equation}
On retained edges far from the origin,
\begin{equation}
 |D_bh|\leq C r_b^{-1/2},\qquad
 |D_bh-P_b|\leq C r_b^{-5/2}.
 \label{eq:core-2}
\end{equation}
Here and below finitely many edges near the tip can be bounded
separately. To prove \eqref{eq:core-2}, the Cauchy--Riemann equations identify $P_b$
with the integral of the appropriate derivative of $h$ over the dual
unit segment. Compare each of the primal and dual integrals with its
midpoint value. Third derivatives of the square root are bounded by
$C r^{-5/2}$. The resulting midpoint remainders give \eqref{eq:core-2}. Segments
adjacent to the cut use the analytic continuation from their own
side; no retained primal or dual segment crosses the cut in its
interior. The number of lattice edges in a shell of radius $r$ and
width one is $O(r)$, so $Dh-P\in\ell^2(\BB_0)$.

\subsection{Renormalized energy}

For $v\in\HH$ define
\begin{align}
 E(v,K)={}&\sum_{b\in\BB_0}
 \bigl\{\phi(KD_bh+D_bv)-\phi(KD_bh)-\tau(KD_bh)D_bv\bigr\}
 \notag\\
 &+\sum_{b\in\BB_0}\bigl\{\tau(KD_bh)-KP_b\bigr\}D_bv
 \notag\\
 &+\sum_{b\in\Gamma}
 \bigl\{\phi(KD_bh+D_bv)-\phi(KD_bh)\bigr\}.
 \label{eq:core-3}
\end{align}
This is a definition by convergent expressions, not an arbitrary
ordering of an infinite total energy.

\begin{lemma}[Well-defined energy and smooth force]
Equation \eqref{eq:core-3} defines a $C^\infty$ function on
$\HH\times(0,\infty)$. For compactly supported $v$ it equals the
ordinary finite energy difference over all physical bonds. Its first
variation is the continuous functional
\begin{align}
 F(v,K)[w]={}&\sum_{b\in\BB_0}
       \bigl\{\tau(KD_bh+D_bv)-KP_b\bigr\}D_bw
 \notag\\
 &+\sum_{b\in\Gamma}\tau(KD_bh+D_bv)D_bw.
 \label{eq:core-4}
\end{align}
Thus $F(v,K)=0$ is equivalent to the physical nearest-neighbor force
balance at every vertex, with no interaction removed.
\end{lemma}
\begin{proof}
The elementary bounds $|\phi''|\leq1$ and
$|\tau(t)-t|\leq3|t|^3$, together with \eqref{eq:core-2}, imply
\[
 \bigl(\tau(KDh)-KP\bigr)\in\ell^2(\BB_0).
\]
The first sum in \eqref{eq:core-3} is bounded absolutely by
$\frac12\norm v_{\HH}^2$, and the second converges by
Cauchy--Schwarz. Also
$\tau(KDh+Dv)-\tau(KDh)\in\ell^2(\BB_0)$.

The predictor opening at $c_n$ has square between $4n+2$ and $4n+3$.
The trace lemma implies that, at every fixed $(v_0,K_0)$ with $K_0>0$,
there are a neighborhood and constants $a>0,N$ such that
\begin{equation}
 |KD_{c_n}h+D_{c_n}v|\geq a\sqrt{n+1}\qquad(n\geq N)
 \label{eq:core-5}
\end{equation}
uniformly in that neighborhood. Every positive-order derivative of
$\phi$ is a polynomial times $e^{-3t^2}$. Thus the last sum in \eqref{eq:core-3},
its first variation, and all higher derivatives converge locally
uniformly, even after any fixed number of trace factors
$O(\sqrt{n+1})$ is inserted.

For completeness, the retained-edge differentiability of the force
also follows in operator norm. Every derivative of $F$ containing
at least one $v$ derivative has at least two $\ell^2$ differences,
one from a variation and one from its test argument; all additional
factors are bounded in $\ell^\infty$. For derivatives of $F$ only in $K$,
the first differentiated force coefficient is
\[
 \phi''(KDh+Dv)Dh-P=(Dh-P)+(\phi''(KDh+Dv)-1)Dh.
\]
It belongs to $\ell^2$: use the global Lipschitz bound for $\phi''$
to handle $Dv$, and $\phi''(t)-1=O(t^2)$ for the predictor part.
The second $K$ derivative is controlled by
$C(|Dh|^3+|Dv||Dh|^2)$, because $|\phi'''(t)|\leq18|t|$.
For three or more $K$ derivatives, $|Dh|^j\in\ell^2$ for $j\geq3$.
These bounds and the bounded derivatives of the Gaussian give local
operator-norm continuity of every derivative of $F$. They justify
termwise differentiation and establish that $F$ is jointly smooth.
Since $E(0,K)=0$ and the already justified first variation is $F$,
\[
 E(v,K)=\int_0^1F(tv,K)[v]\,\dd t.
\]
Differentiation of this integral establishes the asserted joint
smoothness of $E$, without any separate assumption on a
conditionally convergent scalar derivative.

For compactly supported functions, \eqref{eq:core-1} removes the reference flux
from \eqref{eq:core-3} and \eqref{eq:core-4}. Summation by parts then gives
\begin{equation}
 F(v,K)[w]=\sum_{x\in\Lambda}
 \left(\sum_{y\sim x}\tau(U(x)-U(y))\right)w(x),
 \qquad U=Kh+v,
 \label{eq:core-6}
\end{equation}
when $w$ is compactly supported. If $F=0$, testing individual vertex
variations gives force balance. Conversely those balances make
\eqref{eq:core-6} zero for every compact test; density and continuity imply
$F=0$ on $\HH$.
\end{proof}

The homogeneous and nonlinear corrections used in candidate fields,
\begin{equation}
 \eta=r^{-1/2}\sin(\theta/2),\qquad
 \chi=r^{-1/2}\left\{\frac9{32}\log r\sin(\theta/2)
                    +\frac3{64}\sin(5\theta/2)\right\},
 \label{eq:core-7}
\end{equation}
belong to $\HH$ after arbitrary finite-core definitions. Their
retained gradients are respectively $O(r^{-3/2})$ and
$O((\log r)r^{-3/2})$, and radial cutoffs converge in $\HH$.
A coefficient multiplying $\eta$ is therefore a candidate-construction
parameter, not an additional independent unknown when the correction
already ranges over all of $\HH$.

\begin{lemma}[Compact-cutoff scalar identities]
Suppose $Q\in\HH$ agrees outside a finite core with $b\eta$, and a
continuous functional $G\in\HH^*$ has nodal coefficients $g_x$ on
compactly supported tests. If $\sum_x|g_xQ(x)|<\infty$, then
\begin{equation}
 G[Q]=\sum_xg_xQ(x).
 \label{eq:core-8}
\end{equation}
This applies to the force, tangent, and load scalar sums whenever
their stated nodal tail bounds make the displayed series absolutely
convergent. For odd $g$ and odd $Q$, the right side is twice the
upper-half-plane sum.
\end{lemma}
\begin{proof}
Let $\vartheta_R$ be a radial cutoff equal to one inside radius $R$,
zero outside $2R$, with $|D\vartheta_R|\leq C/R$. Then
$\vartheta_RQ\to Q$ in $\HH$. Indeed the gradient-tail term tends
to zero and the cutoff term is bounded in squared norm by
$C R^{-2}\sum_{R\leq r\leq2R}r^{-1}=O(R^{-1})$.
Continuity gives $G[Q]=\lim_R G[\vartheta_RQ]$; absolute convergence
and dominated convergence give \eqref{eq:core-8}. The parity statement follows by
pairing reflected vertices. No assertion about conditionally summed
raw energies is used.
\end{proof}

\section{Hessian structure and the full-space stability index}

For $U=Kh+v$, the Hessian is the bounded symmetric form
\begin{equation}
 A_U[z,w]=\sum_{b\in\BB}\phi''(D_bU)D_bzD_bw.
 \label{eq:core-9}
\end{equation}
Identify $\HH^*$ with $\HH$ by the reference inner product.

\begin{lemma}[Compact perturbation of the reference operator]
For every $v\in\HH$ and $K>0$, $A_U=I+C_U$ with $C_U$ compact and
self-adjoint. Consequently $A_U$ is Fredholm of index zero.
\end{lemma}
\begin{proof}
On retained bonds, $D_bv\to0$ outside finite sets since $Dv\in\ell^2$,
and $D_bh\to0$ by \eqref{eq:core-2}. Thus $\phi''(D_bU)-1\to0$.
Multiplication by this sequence on $\ell^2(\BB_0)$ is the
operator-norm limit of finite-rank diagonal multipliers. Composing
with the isometric difference map and its adjoint preserves
compactness. On deleted bonds, \eqref{eq:core-5} gives
\[
 \sum_n|\phi''(D_{c_n}U)|\norm{D_{c_n}}^2<\infty.
\]
Their Hessian is an operator-norm convergent sum of rank-one forms.
This proves compactness and symmetry. To see the Fredholm assertion
directly, approximate $C_U$ in norm by a finite-rank operator with
error less than one half. On the orthogonal complement of a suitable
finite-dimensional subspace the corresponding diagonal block of
$I+C_U$ is invertible by a Neumann series. Its Schur complement is
finite-dimensional. Hence the range is closed and kernel and
cokernel are finite-dimensional; symmetry makes their dimensions
equal.
\end{proof}

\begin{lemma}[Codimension-one decomposition]
Let $A$ be a bounded symmetric form on $\HH$ and suppose
\begin{equation}
 A[x,x]\geq\alpha\norm x_{\HH}^2\qquad(x\in\XX),\quad\alpha>0.
 \label{eq:core-10}
\end{equation}
There is a unique $y\in\HH$ such that
\[
 \ell(y)=1,\qquad A[y,x]=0\quad(x\in\XX).
\]
Writing $a=A[y,y]$, every $w\in\HH$ has the decomposition
$w=x+\ell(w)y$ with $x\in\XX$, and
\begin{equation}
 A[w,w]=A[x,x]+a\ell(w)^2.                 \label{eq:core-11}
\end{equation}
If $a>0$ the form is coercive on $\HH$; if $a<0$ it has exactly one
negative direction; and if $a=0$ its kernel is exactly
$\operatorname{span}\{y\}$. In the last case it is coercive with
constant at least $\alpha$ on $y^\perp$.
\end{lemma}
\begin{proof}
The positive restriction of $A$ to $\XX$ is invertible. For example,
if its norm is at most $L$, the map $I-tA_{\XX}$ has norm less than
one for sufficiently small positive $t$, by
$\norm{(I-tA)x}^2\leq(1-2\alpha t+L^2t^2)\norm x^2$.
Its Neumann series gives the inverse, whose norm is at most
$1/\alpha$ by \eqref{eq:core-10}. Solve
$A[x_\zeta,x]=A[\zeta,x]$ on $\XX$ and put $y=\zeta-x_\zeta$.
This proves existence, uniqueness, and \eqref{eq:core-11}. The decomposition and
its inverse are bounded, so $a>0$ implies full coercivity. Formula
\eqref{eq:core-11} gives the other index and kernel conclusions. If $w\perp y$
and $x=w-\ell(w)y$, then $\norm x^2=\norm w^2+\ell(w)^2\norm y^2$;
when $a=0$, \eqref{eq:core-10}--\eqref{eq:core-11} yield $A[w,w]\geq\alpha\norm w^2$.
\end{proof}

At an odd physical state, the first variation annihilates even
variations. This follows either from \eqref{eq:core-6} or by reflecting the energy
and using the evenness of $\phi$. Thus an odd-sector equilibrium is
a full-space equilibrium. The Hessian preserves parity. If \eqref{eq:core-10} is
certified in the full space, all even directions are controlled by
it, since their pin difference is zero. The lifting $y$ in the lemma
is odd: reflection with a simultaneous change of sign fixes both its
normalization and its defining equations. Hence the scalar $a$
determines the full-space Morse index, not merely an index within
the odd subspace.

\section{Existence from uniform constrained coercivity}

\begin{proposition}[A projected residual certificate]
Fix a load and a physical opening, and let $\bar U$ be a center with
those values. Suppose that for $w\in\XX$ with $\norm w\leq r$,
\begin{equation}
 A_{\bar U+w}[x,x]\geq\alpha\norm x^2\quad(x\in\XX),
 \qquad \norm{F(\bar U)|_{\XX}}\leq\varepsilon<\alpha r.
 \label{eq:core-12}
\end{equation}
Assume also that the Hessian norms are uniformly bounded on this
ball. In the present model, a uniform Gaussian bound on the slit
coefficients and the bound $|\phi''|\leq1$ on retained bonds suffice.
Then exactly one constrained critical point $U=\bar U+w$ lies in
that closed ball, and
\begin{equation}
 F(U)|_{\XX}=0,\qquad \norm w\leq\varepsilon/\alpha<r.
 \label{eq:core-13}
\end{equation}
For a smooth parameterized family of centers and uniform strict
bounds, the exact constrained state is smooth in the parameters.
\end{proposition}
\begin{proof}
Use the Riesz representative $G(w)$ of $F(\bar U+w)|_{\XX}$.
Let $L$ be a finite uniform Hessian norm on the ball. Such a bound
is supplied by the explicit slit-tail bounds in an application; it
is not inferred merely from continuity on a bounded subset of an
infinite-dimensional space. Consequently $G$ is $L$-Lipschitz and
$\alpha$-strongly monotone there. For sufficiently small $t>0$,
\[
 T(w)=w-tG(w),\qquad
 \norm{T(w)-T(z)}\leq q\norm{w-z},
 \quad q=\sqrt{1-2\alpha t+L^2t^2}<1.
\]
Choose $t$ additionally so that
$\alpha-L^2t/2>\varepsilon/r$. Since
$1-q\geq t\alpha-t^2L^2/2$, we have
$q r+t\varepsilon<r$. Thus $T$ maps the closed ball strictly into
itself and is a contraction. Its iterates form a Cauchy sequence
and give a unique fixed point, which is a zero of $G$.
Strong monotonicity gives
$\alpha\norm w^2\leq-\ip{G(0)}w\leq\varepsilon\norm w$,
proving \eqref{eq:core-13}. At the root the derivative on $\XX$ is invertible by
\eqref{eq:core-10}. The local implicit function theorem, or its usual proof by
parameter-dependent contraction, gives smooth dependence. The
strict radius margin makes this a statement on open neighborhoods,
including at parameter-cell interfaces.
\end{proof}

For odd centers, a full-space application of this proposition gives
an odd root by uniqueness and reflection symmetry. Equivalently,
one may construct the root in the odd sector and use full-space
coercivity for its subsequent uniqueness and stability conclusions.

\begin{lemma}[Stress interpolation]
Let $U_t=(1-t)U_0+tU_1$ be an affine physical family, whose endpoints
may have different loads. Put $\delta_b=D_b(U_1-U_0)$ and let
$m_b$ bound $|\phi'''|$ between the endpoint strains. Then the
force interpolation remainder has stress coefficients bounded by
\begin{equation}
 \frac{t(1-t)}2m_b\delta_b^2\leq\frac18m_b\delta_b^2.
 \label{eq:core-14}
\end{equation}
Its projected dual norm is bounded by the $\ell^2$ norm of these
coefficients on $\BB_0\setminus\{p\}$ plus
$\sum_n\sqrt{2n+2}$ times their absolute values on $c_n$.
For a fixed test $q$, the corresponding scalar bound is their
absolute sum weighted by $|D_bq|$.
\end{lemma}
\begin{proof}
For a scalar $C^2$ function on $[0,1]$, subtract its endpoint linear
interpolant. The Green kernel of the second derivative with zero
endpoint values has integral $t(1-t)/2$ in absolute value. Apply this
to $\tau(D_bU_0+t\delta_b)$, whose second derivative is
$\phi'''(D_bU_t)\delta_b^2$. Pair the resulting stress with a test
vector and use Cauchy--Schwarz and the constrained trace bound.
The result is applicable whenever the stated sums converge.
For the present far fields this convergence must retain the decay
$|\phi'''(t)|\leq18|t|$: if loads differ,
$\delta_b=O(r^{-1/2})$ and the remainder is $O(r^{-3/2})$.
One must not replace this argument by an energy norm of $U_1-U_0$,
which need not belong to $\HH$.
\end{proof}

The same argument can be applied successively to the two coordinates
of a bilinear four-corner family. When $q_t=(1-t)q_0+tq_1$ with
$\ell(q_i)=1$, the scalar interpolation additionally contains
\begin{equation}
 t(1-t)(F(U_0)-F(U_1))[q_1-q_0].
 \label{eq:core-15}
\end{equation}
Since $q_1-q_0\in\XX$, its absolute value is bounded by
$(\varepsilon_0+\varepsilon_1)\norm{q_1-q_0}/4$ using the endpoint
projected residual bounds.

\section{Physical-opening reduction and strict load monotonicity}

For odd configurations use the physical opening $c=U(p_+)$.
A convenient fixed coordinate representation is
\begin{equation}
 U(c,K)=Kh+(c-Kh(p_+))\zeta+w(c,K),\qquad w(c,K)\in\XX_{\mathrm o}.
 \label{eq:core-16}
\end{equation}
Suppose the constrained equation has been validated smoothly on a
parameter region. Define
\begin{equation}
 R(c,K)=F(U(c,K))[\zeta].
 \label{eq:core-17}
\end{equation}
Because $F(U)|_{\XX}=0$, the full equilibrium equation is equivalent
to $R=0$. For an odd state, $R$ equals twice its physical upper pin
force. Define the fixed-load lifting and the fixed-opening load
response by
\[
 y=U_c\in\HH_{\mathrm o},\qquad W=U_K\in h+\HH_{\mathrm o}.
\]
They satisfy
\begin{equation}
 \ell(y)=1,\quad A_U[y,\XX]=0,\qquad
 \ell(W)=0,\quad A_U[W,\XX]=0.
 \label{eq:core-18}
\end{equation}
The pairing involving $W$ uses the renormalized continuous extension
\[
 A_U[h,w]:=\sum_{b\in\BB_0}
   \bigl(\phi''(D_bU)D_bh-P_b\bigr)D_bw
   +\sum_{b\in\Gamma}\phi''(D_bU)D_bhD_bw.
\]
This is $F_K(v,K)[w]$ with $v$ fixed. For $W=h+z$ set
$A_U[W,w]=A_U[h,w]+A_U[z,w]$. Its continuity was proved in the
smooth-force lemma; on compact tests it is the physical linearized
pairing by the divergence-free identity.

\begin{lemma}[Strict physical load sign]
Suppose the fixed-opening Hessian is coercive and every bond lying
entirely within either open half-plane has strictly positive
stiffness. Then
\begin{equation}
 R_K(c,K)<0.                                          \label{eq:core-19}
\end{equation}
The seam stiffnesses need not be positive.
\end{lemma}
\begin{proof}
It suffices to prove positivity of $W$ on the upper half-plane away
from the pin. Let $h_0=h-h(p_+)\zeta$. On a finite exhaustion choose
odd compact corrections $z_N$, vanishing at the pin, so that
$W_N=h_0+z_N$ satisfies $A_U[W_N,x]=0$ for every such test supported
in the finite domain. Coercivity supplies unique solutions. The
right-hand side involving $h_0$ is a continuous dual functional.
The finite-dimensional spaces are dense in $\XX_{\mathrm o}$:
approximate by compact functions, project to odd functions, and
repair their pin value with $\zeta$. The coercive Galerkin estimate
therefore gives $W_N-W\to0$ in the correction energy norm.

Outside the finite domain $W_N=h>0$ on the upper half-plane, while
$W_N(p_+)=0$. Its negative part on the upper half-plane, extended
oddly, is an admissible finite test. Divide the full quadratic form
by two. The resulting form is
\[
 B(f,g)=\sum_{b\text{ in the upper half-plane}}a_bD_bfD_bg
         +2\sum_{j\text{ on the seam}}a_jf_jg_j.
\]
All bulk $a_b$ are positive. For real $s,t$,
$(s-t)(s_- -t_-)\leq-(s_- -t_-)^2$, where
$s_-=\max(-s,0)$. On the seam $W_N(W_N)_-=-(W_N)_-^2$,
regardless of the sign of $a_j$. Thus
\[
 0=B(W_N,(W_N)_-)\leq-B((W_N)_-,(W_N)_-)\leq0,
\]
and coercivity forces $(W_N)_-=0$. Point evaluations in the odd
gauge are continuous, so $W\geq0$ in the upper half-plane.

If $W$ vanished at a nonpin site, its homogeneous equation there
and the positive bulk coefficients would force every upper neighbor
to vanish. The upper graph with the pin removed is connected; hence
$W$ would vanish everywhere. This is impossible because
$W\in h+\HH$ and $h\notin\HH$. Therefore $W>0$ at all nonpin
upper vertices. The derivative of the upper pin force is
\[
 -\sum_{\substack{z\sim p_+\\z_2>0}}
        \phi''(U(z)-U(p_+))W(z)<0.
\]
The seam term is zero since $W(p_+)=0$. Multiplying by two gives
\eqref{eq:core-19}.
\end{proof}

\begin{corollary}[A unique smooth load graph]
Suppose constrained states have been validated for every point of a
closed parameter strip $K_-(c)\leq K\leq K_+(c)$, the hypotheses of
the load-sign lemma hold throughout, and the exact reduced forces
satisfy
\[
 R(c,K_-(c))>0,\qquad R(c,K_+(c))<0
\]
uniformly. Then there is a unique root $K=K(c)$ in the strip for
every opening. The graph is smooth and lies strictly between the
walls.
\end{corollary}
\begin{proof}
Continuity and the intermediate value theorem give a root. Strict
monotonicity in $K$ gives uniqueness. Equation \eqref{eq:core-19} and the implicit
function theorem give smoothness. Shared endpoint seeds and their
strict existence margins identify neighboring parameter-cell
solutions; alternatively the common-class lemma below supplies the
necessary identification.
\end{proof}

\section{Scalar correction, Schur bounds, and simple folds}

Let $\bar U$ be a center and $U=\bar U+w$ its exact constrained
state, with $w\in\XX$ and $\norm w\leq\rho$. Let $q\in\HH_{\mathrm o}$
be an explicit lifting normalized by $\ell(q)=1$. Suppose
\[
 \norm{A_{\bar U}[q,\cdot]|_{\XX}}\leq r_q
\]
and, throughout the correction ball,
\begin{equation}
 |E'''[q,x,x]|\leq B_q\norm x^2\qquad(x\in\XX).
 \label{eq:core-20}
\end{equation}
Then constrained criticality and Taylor's formula give
\begin{equation}
 \left|R-F(\bar U)[q]\right|
 \leq r_q\rho+\tfrac12B_q\rho^2.                \label{eq:core-21}
\end{equation}
Indeed $q-\zeta\in\XX$, so $R=F(U)[q]$, and the linear Taylor
term is $A_{\bar U}[w,q]$. One explicit admissible $B_q$ is
\begin{equation}
 \sup_{b\in\BB_0\setminus\{p\}}m_b|D_bq|
 +\sum_{n\geq0}(2n+2)m_{c_n}|D_{c_n}q|,
 \label{eq:core-22}
\end{equation}
where $m_b$ bounds $|\phi'''|$ on the entire strain tube.

At the exact state let $y$ be its exact lifting. If
\[
 \norm{(A_U-A_{\bar U})[q,\cdot]|_{\XX}}\leq B_q\rho,
 \qquad
 |A_U[q,q]-A_{\bar U}[q,q]|\leq C_{qq}\rho,
\]
then
\begin{align}
 \norm{y-q}&\leq e:=\frac{r_q+B_q\rho}{\alpha},\notag\\
 A_{\bar U}[q,q]-C_{qq}\rho-
       \frac{(r_q+B_q\rho)^2}{\alpha}
 &\leq A_U[y,y]
 \leq A_{\bar U}[q,q]+C_{qq}\rho.              \label{eq:core-23}
\end{align}
To prove this, solve $A_U[e_0,x]=A_U[q,x]$ on $\XX$.
Then $y=q-e_0$, $\norm{e_0}\leq e$, and exactly
\[
 A_U[y,y]=A_U[q,q]-A_U[e_0,e_0],\qquad
 0\leq A_U[e_0,e_0]\leq(r_q+B_q\rho)^2/\alpha.
\]
These estimates distinguish an approximate lifting from a genuine
kernel vector; no small finite-matrix eigenvalue is being used.

For cubic certification, suppose the following bounds hold throughout
the correction tube:
\[
 \norm{E'''[q,q,\cdot]|_{\XX}}\leq C_{qq},\quad
 |E'''[q,x,x]|\leq B_q\norm x^2,\quad
 |E'''[x,x,x]|\leq M_{3,X}\norm x^3,
\]
and $\norm{E''''[q,q,q,\cdot]|_{\XX}}\leq C_{qqq}$.
Expansion with $y-q\in\XX$ gives
\begin{equation}
 |E'''_U[y,y,y]-E'''_{\bar U}[q,q,q]|
 \leq C_{qqq}\rho+3C_{qq}e+3B_qe^2+M_{3,X}e^3.
 \label{eq:core-24}
\end{equation}
Every constant can be bounded by retained stress norms and the crack
trace sums. In particular, \eqref{eq:core-24} need not use a wasteful global bound
proportional to $\norm q^3$.

\begin{lemma}[Reduced derivatives and fold conditions]
For the constrained reduction,
\begin{equation}
 R_c=A_U[y,y],\qquad R_{cc}=E'''_U[y,y,y].       \label{eq:core-25}
\end{equation}
At a point of the equilibrium graph where $R_c=0$, the full Hessian
has kernel $\operatorname{span}\{y\}$ and
\begin{equation}
 F_K[y]=R_K<0,\qquad
 \ip{F_{vv}[y,y]}y=E'''[y,y,y].                \label{eq:core-26}
\end{equation}
If the cubic is nonzero, the point is a simple quadratic fold, with
\begin{equation}
 K'=0,\qquad K''=-E'''[y,y,y]/R_K\neq0.       \label{eq:core-27}
\end{equation}
Its complementary inverse and stability change are those of the
full-space codimension-one decomposition.
\end{lemma}
\begin{proof}
Differentiate \eqref{eq:core-17}, using $y-\zeta\in\XX$ and
$A_U[y,\XX]=0$, to obtain $R_c=A_U[y,y]$. At fixed $K$ the second
opening derivative $y_c$ lies in $\XX$, so differentiating this
identity gives $R_{cc}=E'''[y,y,y]$; the terms containing $y_c$
vanish. If $R_c=0$, the codimension-one lemma gives the kernel.
Writing $W=h+z$ with $\ell(z)=-h(p_+)$ shows
\[
 R_K=A_U[W,y]=F_K[y]-h(p_+)R_c.
\]
Thus \eqref{eq:core-26} holds at a fold. Differentiate $R(c,K(c))=0$ twice at a
point where $R_c=0$ to obtain \eqref{eq:core-27}. The full derivative of $(F,c)$
is invertible there because $A|_{\XX}$ is invertible and $R_K\neq0$;
this is precisely a locally unique quadratic turning point, not a
higher-dimensional bifurcation.
\end{proof}

\begin{lemma}[Exactly one fold in a window]
Let a connected opening interval lie on a validated equilibrium
graph. Suppose the Schur scalar $S(c)=R_c(c,K(c))$ has opposite
strict signs at its endpoints. Suppose further that the cubic in
\eqref{eq:core-25} has one strict sign at every zero of $S$ in this interval,
consistent with the endpoint signs. Then there is exactly one zero
of $S$, and it is a simple fold.
\end{lemma}
\begin{proof}
The intermediate value theorem gives a zero. Along the graph,
\[
 S'(c)=R_{cc}+R_{cK}K'.
\]
At a zero of $S$, $K'=-R_c/R_K=0$, so $S'$ has the nonzero cubic
sign. All zeros therefore cross in the same orientation. There
cannot be two: between two crossings in the same orientation,
continuity would require a crossing in the opposite orientation.
More formally, zeros are isolated by their nonzero derivatives;
an accumulating sequence in the compact interval would have a
zero with zero derivative, including neither endpoint because their
values are nonzero. The resulting finite ordered list cannot have
successive crossings with the same orientation. The fold conclusion
follows from the preceding lemma.
\end{proof}

Thus strict Schur signs outside two disjoint windows, together with
the two window certificates, prove exactly two folds. The sign of
$S$ is the full stability sign. Because $R_K<0$, it is also the sign
of $K'(c)$. A trapping interval follows only after the actual fold
loads are enclosed. If a fold is proved in a strip
$K_{\rm lin}(c)\pm\delta$, the direct safe enclosure is
\[
 [\min_cK_{\rm lin}(c)-\delta,\ \max_cK_{\rm lin}(c)+\delta].
\]
A bracket at one fixed opening alone does not give a tighter
fold-load enclosure.

\section{Identification of overlapping constructions}

\begin{lemma}[Fixed-load common-class uniqueness]
Let $\mathcal C$ be a convex class of physical configurations at a
fixed load. Suppose $A_U|_{\XX}\geq\alpha I$ throughout
$\mathcal C$. Two constrained critical points in $\mathcal C$ with
the same opening are equal modulo constants.
\end{lemma}
\begin{proof}
Their difference $w$ belongs to $\XX$. Subtract the projected force
equations and test with $w$. Integration along the segment in
$\mathcal C$ gives $0\geq\alpha\norm w^2$, hence $w=0$.
\end{proof}

This lemma is a necessary gluing device when a separately validated
fold neighborhood and a global continuation mesh do not share
identical seed balls. One may instead prove that both exact states
belong to a single uniqueness ball. Separate local uniqueness
claims, without an overlap or common-class certificate, are
insufficient.

\begin{lemma}[Common-opening uniqueness, including the load]
Let $\mathcal C$ be convex in the physical $(v,K)$ coordinates, with
loads in a positive interval. Suppose all its half-plane bulk
stiffnesses are positive and $A_U|_{\XX}\geq\alpha I$ throughout
$\mathcal C$. Two odd full equilibria in $\mathcal C$ with the same
physical opening have the same load and displacement.
\end{lemma}
\begin{proof}
For equilibria $U_i=K_i h+v_i$, put
$w=U_2-U_1=\kappa h+z$, with $\ell(w)=0$, and define the secant
coefficients
\[
 a_b=\int_0^1\phi''\bigl(D_b((1-t)U_1+tU_2)\bigr)\,\dd t.
\]
The secant form is coercive on $\XX$ and its half-plane bulk
coefficients are positive. It is bounded on $\HH$, and its pairing
with $h$ is continuous. For retained edges this follows by writing
$(a_b-1)D_bh$ as the sum of an $O(|Dh|^3)$ sequence and terms bounded
by $C(|Dv_1|+|Dv_2|)|Dh|$. These are square summable. On the slit,
the trace lemma is uniform along the compact segment, so its
coefficients have a uniform Gaussian bound.

Subtracting the force equations gives the secant homogeneous
equation for $w$ at every vertex. If $\kappa>0$, $w/\kappa$ is the
fixed-opening affine response in $h+\HH$. The proof of the strict
load-sign lemma applies to this secant operator. Its pin force must
be strictly negative, contradicting the secant force equation at
the pin. If $\kappa<0$, interchange the equilibria. Therefore
$\kappa=0$, and testing with $w\in\XX$ gives $w=0$ by coercivity.
\end{proof}

Fixed strain intervals together with a load interval define such a
convex class. A union of unrelated validation balls generally does
not. Only the particular adjacent or terminal common classes used
in the proof need be certified; no global convexity of the entire
one-cell envelope is implied.

\section{Translation and the smooth periodic join}

Let $(\mathcal Tv)(x)=v(x-e_1)$ on displacement classes.
Direct reindexing gives
\begin{equation}
 \norm{\mathcal Tv}_{\HH}^2
 =\norm v_{\HH}^2+|D_{c_0}v|^2,\qquad
 \norm{\mathcal T^{-1}v}_{\HH}^2
 =\norm v_{\HH}^2-|D_pv|^2.
 \label{eq:core-28}
\end{equation}
Thus $\mathcal T$ is a bounded automorphism with
$\norm{\mathcal T}\leq2$ and $\norm{\mathcal T^{-1}}\leq1$.
The predictor defect $s=\mathcal Th-h$ belongs to $\HH$: away from
finitely many tip edges,
\[
 s=\tfrac12r^{-1/2}\sin(\theta/2)+O(r^{-3/2}),
 \qquad Ds=O(r^{-3/2}),
\]
and the cutoff argument used above proves membership in the
completion. The correction-space translation at load $K$ is
\begin{equation}
 T_Kv=\mathcal Tv+Ks,
 \qquad Kh+T_Kv=\mathcal T(Kh+v).             \label{eq:core-29}
\end{equation}
For compact tests, physical force reindexing is a finite sum, and
hence
\begin{equation}
 F(T_Kv,K)[\mathcal Tw]=F(v,K)[w].            \label{eq:core-30}
\end{equation}
Density extends this identity to every $w\in\HH$. Differentiation
in $v$ gives Hessian congruence, so translation preserves equilibrium,
kernel dimension, and Morse index.

\begin{proposition}[Terminal matching]
Suppose a locally unique odd equilibrium graph is certified on an
opening interval $J$, and let $U_0=U(c_0)$ be a stable point on it.
Suppose a convex terminal class satisfies the hypotheses of the
common-opening uniqueness lemma, and that
\begin{equation}
 c_*=U_0(-\tfrac12,\tfrac12)\in J,
 \qquad \mathcal TU_0\in\mathcal C_{\rm term},
 \qquad U(c_*)\in\mathcal C_{\rm term}.
 \label{eq:core-31}
\end{equation}
Then
\begin{equation}
 U(c_*)=\mathcal TU_0,\qquad K(c_*)=K(c_0).
 \label{eq:core-32}
\end{equation}
\end{proposition}
\begin{proof}
Translation preserves oddness and gives an exact equilibrium at the
original load. Its opening is $c_*$. Apply common-opening uniqueness
to it and $U(c_*)$ inside the terminal class.
\end{proof}

The following gives explicit sufficient containment inequalities.
Suppose the initial exact state at its unknown load is within
$\rho_0$ in $\XX_{\mathrm o}$ of a convex interpolation of its two
initial wall seeds $U_0^-,U_0^+$. Then its translate is within
$2\rho_0$ of the corresponding convex interpolation of the
\emph{exactly translated} wall seeds. A terminal strain box containing
those translated seeds and all relevant terminal wall seeds, enlarged
by the trace bounds for radius $r_C$, contains the translated state
if $2\rho_0<r_C$. It contains the terminal graph states if their
proved correction bounds are less than $r_C$. Furthermore, putting
$d_\pm=U_0^\pm(-1/2,1/2)$, the translated opening lies in
\begin{equation}
 \left[\min(d_-,d_+)-\rho_0/\sqrt2,
       \max(d_-,d_+)+\rho_0/\sqrt2\right].    \label{eq:core-33}
\end{equation}
This interval must be strictly inside the terminal portion of the
covered graph, and its lower endpoint must exceed $c_0$. Formula
\eqref{eq:core-33} follows from the constrained odd trace bound. Including both
translated wall seeds is important: translating only a central seed
and adding $2\rho_0$ would omit the change of its relaxed correction
as the load varies.

Exact translation means applying the source core/exterior rule
\emph{after} replacing $x$ by $x-e_1$. A boundary strip that translates
from source core data must not be replaced by its analytic far-field
approximation. The leading homogeneous coefficient changes by
$K/2$, but this leading term is not the complete translated tail.

\begin{proposition}[Smooth continuation by one lattice cell]
Assume \eqref{eq:core-32}, assume both endpoints are stable, and suppose the arc
$c\in[c_0,c_*]$ is locally unique as a full-space equilibrium arc.
Then it admits a smooth parameter $s\in[0,T]$, $T>0$, which extends
by translation to a smooth equilibrium branch on the real line with
\begin{equation}
 K(s+T)=K(s),\qquad U(s+T,x)=U(s,x-e_1).
 \label{eq:core-34}
\end{equation}
Any simple folds and their full stability indices repeat under this
extension.
\end{proposition}
\begin{proof}
Stability implies $R_c>0$, and $R_K<0$, so $K'(c)>0$ in neighborhoods
of both endpoints. The full Hessian is invertible there; hence load
$K$ is a regular local coordinate for the solution arc. Translation
preserves $K$. Equation \eqref{eq:core-32}, continuity of translation, and local
fixed-load uniqueness therefore identify the terminal load chart
with the translated initial load chart, including all derivatives.

Choose a positive smooth function $v(c)$ on $[c_0,c_*]$ that equals
$1/K'(c)$ in small endpoint neighborhoods. Such a function is
obtained by positive smooth interpolation with a constant in the
interior; the reciprocal is used only where $K'>0$. Define
\[
 s(c)=\int_{c_0}^c\frac{\dd t}{v(t)},\qquad T=s(c_*).
\]
Near each endpoint $\dd K/\dd s=1$. Thus near $s=0$ the load is
$K_0+s$, and near $s=T$ it is $K_0+s-T$. Define the successive arcs
by $U(s+nT)=\mathcal T^nU(s)$ and $K(s+nT)=K(s)$.
The matching load charts show that the left and right definitions
at each join are the same smooth function of the same local load
coordinate. This proves smoothness and \eqref{eq:core-34}. Each finite power of
translation acts on $\HH$ by \eqref{eq:core-28}--\eqref{eq:core-29}, so every extended state
belongs to the specified energy space at its load. Equation \eqref{eq:core-30}
and Hessian congruence preserve equilibria, kernels, and indices;
the fold pairings are preserved as follows. Differentiating
\eqref{eq:core-30} with $v$ fixed gives
\[
 F_K(T_Kv,K)[\mathcal Tq]
   +A_{\mathcal TU}[s,\mathcal Tq]=F_K(v,K)[q],
 \qquad s=\mathcal Th-h.
\]
At a fold the middle term vanishes because $\mathcal Tq$ is a
kernel vector. Reindexing the absolutely convergent cubic gives
\[
 E'''_{\mathcal TU}[\mathcal Tq,\mathcal Tq,\mathcal Tq]
   =E'''_U[q,q,q].
\]
Normalizing either nonzero kernel vector only multiplies these
pairings by nonzero factors.
\end{proof}

\section{Exact normalization of the certificate data}

If a stored lifting $Q$ has a nonzero exact dyadic pin value, set
$q=Q/\ell(Q)$ exactly; do not replace a pin value close to one by
one without scaling the scalar actions and residual norms. If a
corner field has pin discrepancy $\delta$, repair it by
$\delta\zeta$. Exactly seven retained bonds change, and
\begin{equation}
 \norm{F(U+\delta\zeta)-F(U)}_{\HH^*}
 \leq\sqrt{10}|\delta|,
 \qquad
 \norm{A_{U+\delta\zeta}-A_U}\leq8|\delta|,
 \label{eq:core-35}
\end{equation}
using $|\phi''|\leq1$ and $|\phi'''|<4$. Scalar and tangent changes
are obtained by multiplying by the proved lifting norm. Cubic
changes use $|\phi''''|\leq18$ and the seven affected
$|D_bq|^3$ values. Thus a normalization margin can be justified by
explicit inequalities.

The following certificate sections supply the strict inequalities and
common-class containments required by these analytic results.


\section{The residual certificates and their infinite tails}
\label{sec:core-computation}

This section specifies the force, tangent, interpolation, and norm certificates
used below. Their inputs are finite arrays together with explicit infinite
tails; their conclusions are bounds for the infinite physical model.
Write $\ell(v)=D_pv/2$ and $\XX=\ker\ell$. A tip subtraction is harmless on
$\XX$, but every other physical interaction, including every negative-axis
seam interaction, is retained.

\subsection{Exact inputs and the corrected predictor}
Higher-order far-field expansions for antiplane lattice cracks are
studied in \cite{braun2025}. Here the predictor below enters the
residual estimates, which also bound the contribution from the infinite exterior.

Every stored binary64 entry is interpreted as its exact dyadic real value.
For $R=128$, the upper core comprises
$x=-R+1/2,\ldots,R-1/2$ and $y=1/2,\ldots,R-1/2$, with rows indexed first
by $y$ and then by $x$. Both core arrays are extended oddly across $y=0$.
Outside the core the state and lifting field are
\[
 U=Kh+K^3\chi+a\eta,\qquad Q=b\eta,\qquad
 \eta=r^{-1/2}\sin(\theta/2),
\]
\[
 \chi=r^{-1/2}\left(\frac9{32}\log r\sin\frac\theta2+
                    \frac3{64}\sin\frac{5\theta}2\right).
\]
Here $b$ is the stored parameter named \texttt{aprime}. All authoritative
final center, wall, local, and expanded-patch lifting arrays have
$Q(p_+)=1$ exactly, so $\ell(Q)=1$. Historical exploratory arrays are not
substituted for these final inputs.

The base stream functions are
\[
 \Psi_0=r^{1/2}\cos(\theta/2),\qquad
 \Psi_3=-r^{-1/2}\left(\frac9{32}\log r\cos\frac\theta2+
                       \frac3{64}\cos\frac{5\theta}2\right).
\]
The polar-coordinate Laplacian formula gives
\[
 \Delta\chi=3\operatorname{div}(h_x^3,h_y^3)
 =-\frac9{32}r^{-5/2}
       \left(\sin\frac\theta2+\sin\frac{5\theta}2\right),
\]
and direct differentiation gives
\[
 (-\partial_y\Psi_0,\partial_x\Psi_0)=\nabla h,
 \qquad
 (-\partial_y\Psi_3,\partial_x\Psi_3)
       =\nabla\chi-3(h_x^3,h_y^3).
\]
Thus the force and tangent comparison streams are respectively
\[
 \Psi=K\Psi_0+K^3\Psi_3-a r^{-1/2}\cos(\theta/2),
 \qquad \Psi_Q=-b r^{-1/2}\cos(\theta/2).
\]
They vanish on the negative real axis. Their values at the origin are
defined to be zero when evaluating discrete stream differences.

No trigonometric library evaluation is required. For $y\geq0$ set
$r=\sqrt{x^2+y^2}$ and $t=\sqrt{(r+|x|)/2}$. Then $h=t$ for $x<0$ and
$h=y/(2t)$ for $x\geq0$; the real square root uses the opposite choice.
The corrector can be evaluated as
\[
 \chi=\frac h r\left[\frac9{32}\log r+
        \frac3{64}\left(4(x/r)^2+2x/r-1\right)\right].
\]
The corresponding stream expression uses the real square root, the
opposite sign, and the polynomial $4(x/r)^2-2x/r-1$.

\subsection{Exactly equilibrated discrete streams}
On the dual lattice $\mathbb Z^2$, let $G$ be any scalar function.
For a horizontal primal edge define its comparison stress as the lower
dual value minus the upper dual value. For a vertical primal edge take
the right dual value minus the left dual value. The resulting stress
has zero divergence at every primal vertex: the four dual values cancel
algebraically. This remains true for an arbitrary finite-core extension
and for a cutoff applied to the stream.

For the force certificate define the core stream by exact real sums,
starting with $G(-R,0)=0$ and setting
\begin{align*}
 G(x+1,0)-G(x,0)
   &=\tau(2U(x+1/2,1/2))-\mu\mathbf1_{\{x=0\}},\\
 G(x,j+1)-G(x,j)
   &=-\tau\bigl(U(x+1/2,j+1/2)-U(x-1/2,j+1/2)\bigr).
\end{align*}
Here $\mu$ is a specified exact dyadic number and $\tau=\phi'$.
Extend the stream evenly across the axis. For the tangent certificate
replace each traction by $\phi''(DU)DQ$, and replace $\mu$ by a specified
exact dyadic $\nu$. The interval calculation encloses these exact sums;
rounded previously saved stream values are not their definition.
The corresponding horizontal core residuals therefore vanish exactly.

On the square boundary take the mismatch $G-\Psi$, or $G-\Psi_Q$.
Outside the square, linearly interpolate this mismatch at the intersection
of the ray from the origin with the boundary, and multiply it by
\[
 s^{-1/2}\min\{1,\max\{0,(8-s)/4\}\},
 \qquad s=\max(|x|,|y|)/R.
\]
Add this correction to the appropriate base stream. The interpolation
segment and weight at each dual vertex are determined by exact integer
quotient and remainder. Since $R$ is a power of two, the taper is exactly
dyadic. The correction vanishes identically for $s\geq8$.
All core, boundary, and transition edges through $L=8R$ are included in
the finite computation. Every omitted retained edge has midpoint radius
at least $L$, and its dual endpoints lie in the base-stream region.

If $p_e$ denotes this comparison stress, define
\[
 r_e=\tau(D_eU)-\mu\mathbf1_{\{e=p\}}-p_e.
\]
Discrete telescoping gives, for compactly supported $v$,
\[
 (F(U)-\mu D_p)[v]=\sum_e r_eD_ev.
\]
For the negative seam edge with midpoint $(-m-1/2,0)$, a retained detour
has length $2m+3$. Hence the certified dual bound is
\begin{equation}\label{eq:core-dual-bound}
 \|F(U)|_{\XX}\|_{\XX^*}
 \leq\left(\sum_{e\in\BB_0}r_e^2\right)^{1/2}
       +\sum_{e\in\Gamma}\sqrt{2m(e)+3}\,|r_e|.
\end{equation}
The same formula applies to $A_U Q|_{\XX}$ with
$r_e=\phi''(D_eU)D_eQ-\nu\mathbf1_{\{e=p\}}-p_e$.
The bounds extend to the completed energy space by density.
Neither $\mu$ nor $\nu$ requires a separate approximation-accuracy claim:
the corresponding exact multiple of $D_p$ vanishes on $\XX$.
Since $p\in\BB_0$, $\|D_p\|_{\HH^*}\leq1$, and the full tangent norm is
at most the projected bound plus $|\nu|$.

\subsection{Explicit retained-edge tails}
Let an edge midpoint have radius $r\geq N\geq2$ and set
$\rho=r-1/2$ and $L_r=1+\log(r+1/2)$. If $H=D_eh$, $C=D_e\chi$,
and $p_0,p_3$ are the base-stream stresses, midpoint quadrature gives
\begin{align*}
 |H-p_0|&\leq\tfrac1{32}\rho^{-5/2},\\
 |C|&\leq\tfrac{51}{128}L_r\rho^{-3/2},\\
 |C-p_3-3H^3|&\leq\tfrac{663}{2048}L_r\rho^{-7/2}.
\end{align*}
For completeness, the unit-segment midpoint error is bounded by
$\sup|f''|/24$. Third Cartesian derivatives of $h$ are bounded by
$3\rho^{-5/2}/8$. The identities
\[
 \chi=-\tfrac9{32}\log r\,\operatorname{Im}z^{-1/2}
      -\tfrac3{64}\operatorname{Im}(z^{-3/2}\bar z)
\]
bound its first and third derivatives by
\[
 \frac{18\log r+51}{128}r^{-3/2},\qquad
 \frac{270\log r+1413}{512}r^{-7/2},
\]
respectively. Also $|\partial_t^2(h_e^3)|\leq15\rho^{-7/2}/32$.
The two midpoint errors contribute $1413L_r/6144+15/256$;
replacing the midpoint cube by $H^3$ contributes $9/256$.
Their sum is bounded by $663L_r/2048$, proving the displayed estimate.
The phase has
\[
 |D_e\eta-p_\eta|\leq\tfrac5{32}\rho^{-7/2},\qquad
 |D_e\eta|\leq\tfrac12\rho^{-3/2}.
\]

For the actual phase-dependent state put
\begin{align*}
 D&=K^3\tfrac{51}{128}+|a|/2,\\
 A&=K/2+D\frac{1+\log(N+1/2)}{N-1/2},\\
 C_0&=K/32+\tfrac92A^5,\\
 C_1&=\frac{K^3(663/2048)+|a|(5/32)}{N-1/2}+9DA^2.
\end{align*}
Using $|t e^{-3t^2}-t+3t^3|\leq9|t|^5/2$ gives
\[
 |\tau(D_eU)-p_e|\leq(C_0+C_1L_r)\rho^{-5/2}.
\]
There are at most $57n$ positive-orientation edge midpoints in
$n\leq r<n+1$. Covering the midpoints by disjoint unit squares, separately
for the two edge orientations, proves this full-plane count; it includes
both signs of $y$. Put $V=C_0+C_1(2+\log N)$. The squared retained-edge
force tail is bounded by
\begin{equation}\label{eq:force-tail-explicit}
 57(4/3)^5\left\{\frac{V^2}{N^4}
 +\frac{V^2+(2C_1/3)V+2C_1^2/9}{3N^3}\right\}.
\end{equation}
Indeed the remaining series is bounded by its first term plus the
integral of $(C_0+C_1(2+\log t))^2t^{-4}$.

For the tangent, $|\phi''(t)-1|\leq9t^2$ and
$|DU|\leq A\rho^{-1/2}$ imply
\[
 |\phi''(DU)DQ-p_e^Q|\leq C_Q\rho^{-5/2},\qquad
 C_Q=|b|\left[\frac{5/32}{N-1/2}+\frac92A^2\right].
\]
The squared retained-edge tangent tail is at most
\begin{equation}\label{eq:tangent-tail-explicit}
 57(4/3)^5C_Q^2\left[N^{-4}+\frac1{3N^3}\right].
\end{equation}

\subsection{The physical negative-seam tail}
For midpoint $(-s,0)$ with $s\geq R+1/2$, the corrector and nonnegative
phase increase the predictor opening, so $DU\geq2K\sqrt s$.
The retained detour has length $2s+2$. In the verified monotonicity regime
the weighted force and tangent tails are bounded, respectively, by
\[
 2K\sqrt{2+2/s_0}\,\mathcal S(K,s_0),\qquad
 48K^2|b|\sqrt{2+2/s_0}\,\mathcal S(K,s_0),
\]
where
\[
 \mathcal S(K,s_0)=e^{-12K^2s_0}
 \left[\frac{s_0}{1-e^{-12K^2}}
       +\frac{e^{-12K^2}}{(1-e^{-12K^2})^2}\right].
\]
This is the exact sum of $s e^{-12K^2s}$ over $s=s_0,s_0+1,\ldots$.
For the tangent the program verifies $2K\sqrt{s_0}\geq1/\sqrt2$;
it also verifies the applicable force monotonicity condition.
These bounds control the physical seam forces rather than deleting them.

\subsection{Uniform interpolation and lifting norms}
For two candidates let $U_t=(1-t)U_0+tU_1$ and
$Q_t=(1-t)Q_0+tQ_1$. Interpolate their exact streams with the same weights.
On an edge set $d_j=D_eU_j$, $e_j=D_eQ_j$, $\Delta d=d_1-d_0$,
and $\Delta e=e_1-e_0$. The force and tangent interpolation defects are
bounded by
\[
 \frac18 M_3|\Delta d|^2,\qquad
 \frac18\left[M_4|\Delta d|^2\max(|e_0|,|e_1|)
                +2M_3|\Delta d||\Delta e|\right],
\]
where $M_j$ encloses the indicated derivative on the strain hull. These
follow from the second derivative in $t$ and the interpolation remainder.
The identities evaluated by the interval engine are
\[
 \phi'''(d)=18d(2d^2-1)e^{-3d^2},\qquad
 \phi''''(d)=18(-12d^4+12d^2-1)e^{-3d^2}.
\]
Every finite edge through $L=256$ is included; only the tip term may be
omitted for a projected bound. In the exterior $\chi$ has the affine
coefficient $(1-t)K_0^3+tK_1^3$, not $((1-t)K_0+tK_1)^3$.

For the explicit interpolation tails take $N=L$ and define
\begin{align*}
 A_j&=K_j/2+
 \frac{K_j^3(51/128)(1+\log(N+1/2))+|a_j|/2}{N-1/2},\\
 A_*&=\max(A_0,A_1),\\
 B_*&=|K_1-K_0|/2+
 \frac{|K_1^3-K_0^3|(51/128)(1+\log(N+1/2))+|a_1-a_0|/2}{N-1/2},\\
 Q_*&=\max(|b_0|,|b_1|)/2,\qquad E_*=|b_1-b_0|/2,\\
 M_{3,*}&=18A_*\left[1+\frac{2A_*^2}{N-1/2}\right],\\
 M_{4,*}&=18\left[1+\frac{12A_*^2}{N-1/2}
                       +\frac{12A_*^4}{(N-1/2)^2}\right],\\
 C_f&=M_{3,*}B_*^2/8,\qquad
 C_t=(M_{4,*}B_*^2Q_*+2M_{3,*}B_*E_*)/8.
\end{align*}
The edge defects are at most $C_f\rho^{-3/2}$ and
$C_t\rho^{-5/2}$. Their squared full-plane retained tails are at most
\[
 57(4/3)^3C_f^2[N^{-2}+N^{-1}],\qquad
 57(4/3)^5C_t^2[N^{-4}+1/(3N^3)].
\]
For the omitted seam use twice the preceding endpoint-family geometric
bound with $K_{\min}=\min(K_0,K_1)$ and
$b_{\max}=\max(|b_0|,|b_1|)$. Nonnegative phases guarantee the required
lower opening along the segment. These defects are added to endpoint
residuals; they are uniform for $0\leq t\leq1$.

For the lifting norm the finite reference-energy sum is enclosed through
$N=256$. Square midpoint shells contain at most
$16n+4\leq40(n-1/2)$ edges, and
$|DQ|^2\leq b^2(n-1/2)^{-3}/4$. Thus the omitted squared norm is bounded by
\[
 10b^2\left[\frac1{N-1/2}+\frac1{(N-1/2)^2}\right].
\]
Adding this to the finite squared norm and taking an outward square root
gives the stated full reference norm. No negative seam term belongs to
this reference norm.

\subsection{Outward arithmetic and the saved bounds}
The vector engine uses IEEE-754 binary64 basic arithmetic and
\texttt{nextafter}. Each elementary interval operation expands the
rounded result outward by one representable step. It checks binary
radix, 53-bit significands, round-to-nearest, gradual underflow, finite
inputs, and the power-of-two condition on $R$. If $n$ nonnegative terms
are summed, put $\gamma_n=n/(2^{53}-n)$, rounded upward. The sum is
enlarged using a downward-rounded $1-\gamma_n$ denominator and the
explicit allowance $n\,2^{-1074}$, with outward rounding throughout.

Square-root library calls provide guesses only: outward squared-endpoint
comparisons verify each enclosure, expanding it until both inequalities
hold. Logarithms use exact binary range reduction and the atanh series
through degree $41$, with the explicit geometric remainder. Exponentials
on nonpositive arguments use division by $64$, the degree-$18$ alternating
Taylor polynomial with next-term remainder, and six outward squarings.
Arguments below $-40$ use $[0,2^{-57}]$. The needed inequality follows
from $e^{0.7}>2$. Sign restrictions needed by the alternating remainder
are checked. The constant $\log2$ and the scalar tail expressions
involving logarithms, exponentials, and square roots are enclosed by
\texttt{mpmath.iv} at $60$ decimal digits. No undocumented accuracy bound
for a NumPy transcendental function is assumed.

The saved certificates give the following conservative summaries:
\[
\begin{array}{c|cc}
 \text{input family}&\|F(U)|_{\XX}\|&\|A_U Q|_{\XX}\|\\ \hline
 32\text{ final centers}&<0.000964&<0.003528\\
 64\text{ global load walls}&<0.000965&\text{bounded by transfer}\\
 16\text{ local fold states}&<0.000791&<0.002967
\end{array}
\]
All $32$ final lifting norms are less than $3$. Across the $31$ adjacent
center segments, the projected force interpolation defect is less than
$0.000240$, and the projected tangent interpolation defect is less than
$0.00430$. The assembly certificates use the individual saved bounds,
rather than replacing them by these common maxima. Hash-bound final
inputs, interval-engine sources, and the corresponding outputs are
included with the computational certificate. These residual statements
are combined with coercivity and the scalar validation inequalities in
the proof; a finite residual calculation alone is not an existence claim.

\section{Certification of the full constrained Hessian}

All coercivity statements in this section concern the full space
$\XX=\ker\ell$, where $\ell(v)=D_pv/2$, rather than its odd
subspace. For a physical configuration $U$, write
\[
 a_b(U)=\phi''(D_bU)=(1-6(D_bU)^2)e^{-3(D_bU)^2}.
\]
Its Hessian is
\begin{equation}
 A_U[w,w]=\sum_{b\in\BB_0}a_b(U)(D_bw)^2
          +\sum_{n\geq0}a_{c_n}(U)(D_{c_n}w)^2.
 \label{eq:spectral-full-hessian}
\end{equation}
The negative slit terms are retained throughout the certification.

\subsection{Coefficient intervals and a finite Neumann patch}

Let $U_1,\ldots,U_m$ be the explicitly stored physical centers of
one certified class. Convex interpolation of these fields places
each strain between its smallest and largest corner values.
A correction of energy norm at most $r$ enlarges this interval by
$r$ on retained bonds and by $r\sqrt{2n+3}$ on $c_n$. These
conservative bounds apply without a symmetry assumption. Any
additional allowed load change is incorporated explicitly by its
multiple of $D_bh$. Consequently the coefficient intervals hold on
the entire continuous class, not merely at sampled centers.

The interval engine treats stored binary floating-point values as
exact dyadic rationals. Square roots are enclosed by integer
square-root calculations. Exponentials are enclosed by rational
Taylor inequalities; in particular, the reciprocal of a positive
Taylor lower bound for $e^x$ is an upper bound for $e^{-x}$.
The monotonicity intervals of
$(1-6t^2)e^{-3t^2}$, with its critical points included, give a
rational lower bound $\underline a_b$ for every coefficient.
Finally these bounds are rounded downward to the common
denominator $10^8$. Thus rounding cannot strengthen the claimed
lower quadratic-form bound.

The patch has vertex set
\[
 P_M=\{i+\tfrac12:-10\leq i\leq7\}
       \times\{j+\tfrac12:-M\leq j\leq M-1\}.
\]
Only bonds whose two endpoints belong to $P_M$ enter its matrix.
No artificial exterior Dirichlet condition is imposed. For
$w\in\XX$, choose the constant representative with
$w(p_-)=w(p_+)=0$; this is possible precisely because $D_pw=0$.
Deleting these two coordinates is therefore the physical opening
constraint together with the quotient gauge, not an extra
constraint on admissible perturbations.

Let $\alpha>0$ be the proposed coercivity constant and let
$\sigma>0$ be the recorded tail reserve. On the remaining patch
coordinates form the rational quadratic form
\begin{equation}
 Q_P(z)=
 \sum_{\substack{b\in\BB_0\\b\subset P_M}}
       (\underline a_b-\alpha-\sigma)(D_bz)^2
 +\sum_{\substack{c_n\subset P_M}}
       \underline a_{c_n}(D_{c_n}z)^2.
 \label{eq:spectral-patch-form}
\end{equation}
The basic checks use $M=2$, giving seventy free coordinates.
The stronger common-class checks use taller patches, including
$M=8$. Every omitted retained bond is separately certified to have
$a_b\geq\alpha+\sigma$.

For an omitted slit bond put
$\beta_n=\max(0,-\underline a_{c_n})$. The checker proves
\begin{equation}
 \sum_{c_n\not\subset P_M}(2n+3)\beta_n<\sigma.
 \label{eq:spectral-tail-reserve}
\end{equation}
If $Q_P$ is nonnegative,
\eqref{eq:spectral-full-hessian}--\eqref{eq:spectral-tail-reserve}
and the crack trace inequality give, for every $w\in\XX$,
\begin{align*}
 A_U[w,w]-(\alpha+\sigma)\norm w_{\HH}^2
 &\geq Q_P(w|_{P_M})
      -\sum_{c_n\not\subset P_M}\beta_n|D_{c_n}w|^2\\
 &\geq-\sigma\norm w_{\HH}^2.
\end{align*}
Hence $A_U[w,w]\geq\alpha\norm w_{\HH}^2$ on the full constrained
space. This is a lower bound for the infinite Hessian, not an
inference from a finite eigenvalue calculation.

\subsection{An exact integer positivity witness}

After clearing the common positive denominator in
\eqref{eq:spectral-patch-form}, let $\mathsf A$ be its symmetric
integer matrix. Floating-point Cholesky calculations are used only
to propose a lower triangular integer matrix $\mathsf B$. The
acceptance test uses exact integer arithmetic and verifies that
every diagonal entry of $\mathsf B$ is nonzero and that
\begin{equation}
 \mathsf C=\mathsf B\mathsf A\mathsf B^{\mathsf T},
 \qquad
 \mathsf C_{ii}>\sum_{j\ne i}|\mathsf C_{ij}|
 \quad\hbox{for every }i.
 \label{eq:spectral-integer-witness}
\end{equation}
For every nonzero real vector $z$,
\[
 z^{\mathsf T}\mathsf C z
 \geq\sum_i
 \left(\mathsf C_{ii}-\sum_{j\ne i}|\mathsf C_{ij}|\right)z_i^2
 >0.
\]
The triangular matrix $\mathsf B$ is invertible, so congruence
implies that $\mathsf A$ is positive definite. No accuracy property
of the floating-point Cholesky calculation is an assumption of this
argument: a poorly chosen witness simply fails
\eqref{eq:spectral-integer-witness}.

The product is computed with exact zeros omitted using only the
proved bandwidth of $\mathsf A$ and triangularity of $\mathsf B$.
Thus the acceptance inequalities are finite integer certificates.

\subsection{The omitted infinite exterior}

Outside the stored core, the physical centers use
$Kh+K^3\chi+a\eta$ with the explicit functions defined earlier.
Their derivatives give
\[
 |\chi|\leq
 \frac{(9/32)\log r+3/64}{\sqrt r},\qquad
 |\nabla\chi|\leq
 \frac{(9/32)(\log r+1)+9/64}{r^{3/2}},
 \qquad
 |\nabla\eta|=\frac1{2r^{3/2}}.
\]
These estimates, the recorded parameter intervals, and the
correction radius give the exterior retained-bond lower bounds
used above. The remaining finite annulus is checked bond by bond.

For the ordinary unshifted classes, the slit estimates give
$|D_{c_n}U|\geq(17/20)\sqrt{n+1/2}$ beyond the recorded exterior
radius. Since the negative part of $\phi''$ decreases in magnitude
after $|t|=1/\sqrt2$, the checker may use
\begin{equation}
 \beta_n\leq3(n+\tfrac12)4^{-n}.
 \label{eq:spectral-gaussian-tail}
\end{equation}
Therefore the infinite remainder in
\eqref{eq:spectral-tail-reserve} is bounded by the exact rational
geometric sum
\[
 3\sum_{n\geq N}(2n+3)(n+\tfrac12)4^{-n}.
\]
The finitely many preceding terms are enclosed individually.
All geometric sums are evaluated rationally, and any aggregation
rounding is upward.

Translated centers are evaluated by applying the source
core/exterior definition after replacing $x$ by $x-e_1$.
In particular, a translated strip coming from source core data is
not replaced by its asymptotic formula. The shifted slit indices
and the remaining analytic tail are bounded separately. The
coefficient $a+K/2$ describes only the leading translated
homogeneous term.

\subsection{Accepted classes and replay}

The accepted four-wall corner checks for the thirty-one global
parameter cells give
\[
\begin{array}{c|c}
 \text{cell indices}&\alpha\\ \hline
 0,1&1/5\\
 2,\ldots,26&1/4\\
 27,\ldots,30&1/5
\end{array}
\]
at correction radius $1/100$. The corners themselves span the
load strip, so these checks require no additional load deviation.

The stronger checks, which also identify the local fold
constructions with the global graph, give
\[
\begin{array}{l|c}
 \text{class or opening interval}&\alpha\\ \hline
 \text{upper common class }[0.29,0.32]&0.35\\
 \text{lower common class }[0.40,0.44]&0.29\\
 \text{lower left common class through the fold center}&0.30\\
 \text{upper local intervals }[0.29,0.30],[0.30,0.31],[0.31,0.32]
       &0.35\\
 \text{lower local intervals }[0.40,0.41],[0.41,0.42]&0.32\\
 \text{lower local interval }[0.42,0.43]&0.30\\
 \text{lower local interval }[0.43,0.44]&0.29
\end{array}
\]
These classes have radius $1/100$. The common classes include the
relevant local physical corners and global wall centers, with the
additional recorded load allowance $1/2000$; the local rectangle
corners already span their loads.

The terminal certificate has radius $1/50$ and coercivity
constant $3/20$. Its class contains the exact translates of both
initial load-wall fields and all terminal wall fields at indices
$28$ through $31$, covering openings from $0.47$ to $0.485$.
This is the terminal class used for matching; a smaller hull of
central or terminal seeds is not substituted for it.

The archived \texttt{coercivity\_manifest.json} records the exact
input hashes, field lists, radii, load allowances, patch sizes,
reserves, and accepted constants. The archived
\texttt{replay\_coercivity.py} invokes
\texttt{coefficient\_certificate.py} and
\texttt{patch\_certificate.py} to recompute the coefficient
enclosures, omitted-tail bounds, and integer tests
\eqref{eq:spectral-integer-witness}.
The authoritative terminal scope and its acceptance inequalities
are recorded separately in
\texttt{terminal\_box\_certificate.json}.
These checks certify continuous coefficient classes and the
infinite quadratic form; they do not assume equilibrium, the
existence of folds, or continuation of a numerical branch.


\section{Validation of the two folds and their separation}
\label{sec:local-fold-validation}

We use the full-lattice energy convention throughout. Thus an edge in the
open upper half-plane has weight two, while a vertical edge crossing the
horizontal axis has weight one. No physical interaction is omitted.
Continue to use $\ell(w)=D_pw/2$ and $X=\XX=\ker\ell$;
on odd representatives $\ell(w)=w(p_+)$. The norm is that of $\HH$.
The notation $\mathcal E^{(j)}(U)$ means the fixed-load $j$th
variation of $E(v,K)$ at $v=U-Kh$. For the physical
seam edge at $x=1/2-j$, $j\geq1$, the two horizontal paths to the pinned
sites give
\begin{equation}\label{eq:local-cut-trace}
 |D_jw|\leq\sqrt{2j}\,\|w\|,\qquad w\in X.
\end{equation}
The tip edge has zero difference on $X$.

\subsection{Quantities bounded by the finite certificates}

At each of the explicitly stored candidate fields $\bar U$, let
$\bar A=\mathcal E''(\bar U)$ and choose a field $q$ with $\ell(q)=1$.
In the stored core every number is its exact binary rational value.
Outside that core the fields are defined, rather than approximated, by
\begin{equation}\label{eq:local-exact-fields}
 \bar U=Kh+K^3\chi+a\eta,\qquad q=b\eta,
\end{equation}
where
\[
 h=r^{1/2}\sin(\theta/2),\quad
 \eta=r^{-1/2}\sin(\theta/2),\quad
 \chi=\frac9{32}r^{-1/2}\log r\sin(\theta/2)
      +\frac3{64}r^{-1/2}\sin(5\theta/2).
\]
Every stored lifting is divided by its exact stored tip value. At the tip
we impose the prescribed physical opening exactly. The necessary finite
pin repair has magnitude less than $10^{-12}$ and energy norm less than
$\sqrt{10}\,10^{-12}$. The interval assemblies below include an additional
$10^{-8}$ allowance for this repair and for the corresponding scalar and
operator perturbations. On the expanded grid the opening is the exact
rational $j/200$ whenever its stored metadata differs from that rational
by less than $10^{-12}$; elsewhere the opening metadata is an exact
binary rational. In particular, adjacent expanded patches have exactly
identical endpoint openings.

Here are the localized multilinear bounds used in the proof. For each
edge $e$, let $g_{3,e}$ and $g_{4,e}$ bound respectively
$|\phi'''|$ and $|\phi''''|$ throughout the energy tube of radius $1/100$
around the candidate convex hull. Let $Q_e$ bound $|D_eq|$ for every
lifting in the selected convex hull. In the following maxima and sums,
``retained'' excludes the distinguished tip edge, and ``cut'' means the
negative-axis seams. Define
\begin{align*}
 B&=\max_{\rm retained}g_3Q+\sum_{j\geq1}g_{3,j}Q_j(2j),\\
 M&=\max_{\rm retained}g_3+\sum_{j\geq1}g_{3,j}(2j)^{3/2},\\
 C_2&=\left(\sum_{\rm retained}g_3^2Q^4\right)^{1/2}
       +\sum_{j\geq1}g_{3,j}Q_j^2\sqrt{2j},\\
 C_4&=\left(\sum_{\rm retained}g_4^2Q^6\right)^{1/2}
       +\sum_{j\geq1}g_{4,j}Q_j^3\sqrt{2j}.
\end{align*}
All sums have the full-lattice weights just stated. Cauchy--Schwarz,
$\|D w\|_{\ell^\infty}\leq\|w\|$ on retained edges, and
\eqref{eq:local-cut-trace} prove, for $x,z,t\in X$,
\begin{align}
 |\mathcal E'''[q,x,z]|&\leq B\|x\|\|z\|,&
 |\mathcal E'''[x,z,t]|&\leq M\|x\|\|z\|\|t\|,\nonumber\\
 |\mathcal E'''[q,q,x]|&\leq C_2\|x\|,&
 |\mathcal E''''[q,q,q,x]|&\leq C_4\|x\|.\label{eq:local-multilinear}
\end{align}
These estimates apply throughout the stated tube.

\subsection{Interpolation and correction estimates}

We interpolate the four corner fields of each opening--load patch
bilinearly, always preserving the physical opening and load. Let
$U_t=(1-t)U_0+tU_1$ and $q_t=(1-t)q_0+tq_1$ along one coordinate.
Taylor's theorem bounds the force interpolation stress on each edge by
\[
 \frac18g_3|D(U_1-U_0)|^2.
\]
For the tangent interpolation stress the bound is
\begin{equation}\label{eq:local-tangent-interpolation}
 \frac18g_4|D(U_1-U_0)|^2\max(|Dq_0|,|Dq_1|)
 +\frac14g_3|D(U_1-U_0)|\,|D(q_1-q_0)|.
\end{equation}
Indeed, writing $A_t=\mathcal E''(U_t)$, the exact identity is
\[
 A_tq_t-\big((1-t)A_0q_0+tA_1q_1\big)
 =\big(A_t-(1-t)A_0-tA_1\big)q_t
  -t(1-t)(A_1-A_0)(q_1-q_0).
\]
Apply these one-coordinate estimates successively in the two coordinates.
The sum of the resulting two errors bounds the full bilinear error.
The retained stresses are summed in $\ell^2$; the cut stresses use
\eqref{eq:local-cut-trace}.

For a load wall, pair the force interpolation stress directly with $q_t$.
The remaining scalar cross term is
$t(1-t)\langle F_0-F_1,q_1-q_0\rangle$, hence is bounded by
\[
 \frac14(\varepsilon_0+\varepsilon_1)\|q_1-q_0\|,
\]
where $\varepsilon_i=\|F_i|_X\|$. This uses $q_1-q_0\in X$ and therefore does
not involve the possibly nonzero tip reaction. The identical argument
for the Schur tests along a fixed-opening load edge adds
\[
 \sqrt{3/2}\,\varepsilon_q\max(\|q_0\|,\|q_1\|)
 +\frac14(r_0+r_1)\|q_1-q_0\|,
\]
where $\varepsilon_q$ bounds \eqref{eq:local-tangent-interpolation} and
$r_i=\|(A_iq_i)|_X\|$.
The factor $\sqrt{3/2}$ is needed because $q_t$ need not lie in $X$.
The full cut-edge trace is $\sqrt{2j+1}\,\|q_t\|$, and
$\sqrt{2j+1}\leq\sqrt{3/2}\sqrt{2j}$ for $j\geq1$.
The omitted tip stress is exactly zero on a fixed-opening load edge,
since both its physical strain and its normalized lifting difference
are constant there. Thus this factor converts the computed projected
stress bound into the required full dual bound. No such conversion is
needed when testing on $X$.

The independent coercivity certificates give $A|_X\geq\alpha I$
throughout the radius-$1/100$ candidate tube. Let $\varepsilon$ bound the
interpolated projected force, including all tails, and suppose
$\rho=\varepsilon/\alpha<9/1000$. Minimize the restricted energy on the closed
radius-$9/1000$ ball in $X$. Strong convexity gives weak lower
semicontinuity, and weak compactness gives a minimizer. On the boundary
the outward radial derivative is at least
$\alpha(9/1000)^2-\varepsilon(9/1000)>0$, so the minimizer is interior.
It is the unique constrained equilibrium $U=\bar U+w$ in this tube and
\begin{equation}\label{eq:local-rho}
 \|w\|\leq\rho.
\end{equation}
The inverse of $A|_X$ gives smooth parameter dependence, and uniqueness
glues overlaps.

Suppose $\|\bar A[q,\cdot]|_X\|\leq r$. Let $y$ be the exact normalized
lifting at $U$, determined by $\ell(y)=1$ and $A[y,X]=0$. Set
\[
 r_t=r+B\rho,\qquad e=r_t/\alpha.
\]
Then \eqref{eq:local-multilinear} and the inverse bound give
$\|y-q\|\leq e$. Taylor's theorem gives the scalar correction bound
\begin{equation}\label{eq:local-scalar-correction}
 |F(U)[q]-F(\bar U)[q]|\leq r\rho+\tfrac12B\rho^2.
\end{equation}
The exact reduced Schur coefficient $s=A[y,y]$ obeys
\begin{equation}\label{eq:local-schur-correction}
 \bar A[q,q]-C_2\rho-\frac{r_t^2}{\alpha}
 \leq s\leq \bar A[q,q]+C_2\rho.
\end{equation}
For the lower bound use $s=A[q,q]-A[q-y,q-y]$ and
$A[q-y,q-y]\leq r_t^2/\alpha$. Finally, the actual cubic coefficient
$\gamma=\mathcal E'''(U)[y,y,y]$ satisfies
\begin{equation}\label{eq:local-cubic-correction}
 |\gamma-\mathcal E'''(\bar U)[q,q,q]|
 \leq C_4\rho+3C_2e+3Be^2+Me^3.
\end{equation}
This follows by first changing the field from $\bar U$ to $U$ and then
expanding $(q+(y-q))^3$. These are estimates for the infinite problem,
not conclusions drawn from a truncated eigenvalue calculation.

\subsection{Tail and arithmetic verification}

The finite scalar checks use site sums on an upper square and odd
reflection. In particular the force and Schur sums are respectively
$2\sum_i F_iq_i$ and $2\sum_i(Aq)_iq_i$. The cubic is the sum of the four
neighbor contributions over upper sites, with no additional factor of
two. In the local parameter box
$9/20\leq K\leq1/2$, $0\leq a\leq1/3$, and $0\leq b\leq11/10$,
the exact predictor \eqref{eq:local-exact-fields} has, at radius $r\geq512$,
\[
 |F_i|\leq(1+\tfrac1{10}\log r)r^{-7/2},\quad
 |(Aq)_i|\leq2r^{-7/2},\quad |q_i|\leq\tfrac{11}{10}r^{-1/2}.
\]
For a square cutoff $L$ and $A=L-1/2$, the omitted absolute scalar tails
are consequently at most
\begin{equation}\label{eq:local-scalar-tails}
 \frac{4.62+0.44\log(\sqrt2A)}{A^2},\qquad
 \frac{8.8}{A^2},\qquad \frac{4.1}{A^3}
\end{equation}
for force, Schur, and cubic, respectively. These decimals denote exact
rationals. The point-force estimate uses the exact identity
$\Delta\chi=9\sum_{j=1}^2h_{e_j}^2h_{e_je_j}$, cancellation of the
continuum cubic term, and fourth-order central-difference remainders.
The derivative comparisons reduce to rational inequalities at stencil
radius $511$; $\log511<6237/1000$ follows from
$\sum_{k=0}^{30}(6237/1000)^k/k!>511$. The physical negative seams are
restored in these estimates and have exponentially small, explicitly
bounded forces.

For coefficient and interpolation sums the finite cutoff is $256$.
Square midpoint shells have at most $20n$ edges; the checks enlarge this
to $32n$ and start the integrals at $254$. The retained-tail bounds are
$|DU|\leq r^{-1/2}$, $|Dq|\leq r^{-3/2}$,
$|\phi'''|<2$ throughout the tube, and $|\phi''''|<60$.
Thus the added squared tails for $C_2$ and $C_4$ are
$32/254^4$ and $115200/(7\cdot254^7)$.
For an omitted negative seam $s\geq256.5$, every coefficient or
interpolation summand is at most $100000s^3e^{-2s}$. Its sum is less
than $150000\cdot257^3\,2^{-513}<3\cdot10^{-141}$; the checks add
$10^{-130}$. This also bounds the seams after the full tube perturbation.

The interval implementation rounds every basic operation outward. A
square-root guess is accepted only after outward multiplication verifies
both inequalities. Exponentials are enclosed by a Taylor remainder and
scaling; logarithms by the atanh series and its geometric remainder.
Finite sums use an explicit rounding-error allowance including
subnormals. Extrema of $\phi'''$ and $\phi''''$ are enclosed using their
explicit critical points. The finite candidate arrays are data specifying
\eqref{eq:local-exact-fields}; their numerical construction is not an
assumption of the proof. The accompanying local manifest specifies the
exact arrays, checker sources, and fifty-five replay commands, including
all finite scalar, interpolation, and cubic checks and the two final
assemblies. The force, tangent, and coercivity certificates are checked
independently. Archive paths are relative, so extraction does not require
the original machine or directory.

\subsection{Results of the inequalities}

The four narrow patches use the conservative value $\alpha=1/4$.
The following entries are rational outward bounds on the assembled
infinite-problem quantities. The wall column is a lower bound for both
the positive scalar force on the lower load wall and its negative on the
upper load wall.
\[
\begin{array}{c|c|c|c|c}
 \text{patch}&\rho< & e< &\gamma\ \text{belongs to}&\text{wall margin}>\\\hline
 \text{upper left}&.001903&.067074&[-7.357,-2.384]&.000214\\
 \text{upper right}&.001948&.067012&[-6.888,-2.099]&.000215\\
 \text{lower left}&.003312&.083469&[.778,5.407]&.0000454\\
 \text{lower right}&.003435&.084979&[.811,5.508]&.0000317
\end{array}
\]
At the two outer endpoints of the upper pair the exact Schur coefficient
belongs respectively to $[.0063,.0410]$ and $[-.0401,-.0062]$.
For the lower pair the corresponding intervals are
\[
 [-.0542,-.0099]\quad\text{and}\quad[.0076,.0541].
\]
The exact outer opening endpoints are
\[
 \begin{array}{c|cc}
 &\text{left}&\text{right}\\\hline
 \text{upper}&\frac{2695946955303457}{9007199254740992}
 &\frac{2786018947850867}{9007199254740992}\\
 \text{lower}&\frac{929362559391911}{2251799813685248}
 &\frac{60899909729101}{140737488355328}.
 \end{array}
\]

The wall signs, continuity, and strict scalar loading monotonicity give a
unique equilibrium load at every opening in each patch. At a zero of the
Schur coefficient, its derivative along this graph is exactly $\gamma$:
the load derivative vanishes there, and terms differentiating the
normalized lifting vanish by $A[y,X]=0$. Thus every upper zero is a
strict positive-to-negative crossing, and every lower zero is a strict
negative-to-positive crossing. A continuously differentiable real
function cannot have two zeros all of which have the same strict
crossing orientation: after one such crossing, a second would require
a crossing of the opposite orientation, by the intermediate value
theorem and the local signs at a simple zero. The endpoint signs above
therefore prove exactly one simple fold in each narrow pair. Together
with strict load transversality, the cubic signs give a local maximum
of the load at the upper fold and a local minimum at the lower fold.

To exclude other folds within the larger zones left by the global
stability verification, the same inequalities were checked on seven
patches with exact rational endpoints. The following table gives
outward bounds; every entry in the last column bounds both load walls.
\[
\begin{array}{c|c|c|c|c}
 \text{opening interval}&\alpha&\rho<&\gamma\ \text{belongs to}
 &\text{wall margin}>\\\hline
 \lbrack.29,.30\rbrack&.35&.001558&[-8.063,-2.500]&.0000099\\
 \lbrack.30,.31\rbrack&.35&.001598&[-7.132,-2.107]&.0000425\\
 \lbrack.31,.32\rbrack&.35&.001646&[-6.105,-1.489]&.0000710\\
 \lbrack.40,.41\rbrack&.32&.002471&[1.094,4.523]&.0000757\\
 \lbrack.41,.42\rbrack&.32&.002562&[1.342,4.759]&.0000627\\
 \lbrack.42,.43\rbrack&.30&.002834&[1.301,5.010]&.0000466\\
 \lbrack.43,.44\rbrack&.29&.003041&[1.140,5.093]&.0000314
\end{array}
\]
Here all displayed opening decimals are exact rationals. The common
coercivity classes contain both these corner fields and the global
branch fields. Their uniqueness lemma identifies the resulting graphs.
The same strict-crossing argument excludes further zeros throughout
$[29/100,8/25]$ and $[2/5,11/25]$. The two narrow folds are consequently
the only folds in those zones.

Finally, compare the exact binary rational load of every narrow-patch
corner with the following rational endpoints. Every upper corner load
lies strictly between $46219/100000$ and $46242/100000$, and every lower
corner load lies strictly between $45897/100000$ and $45923/100000$.
An interpolated wall load is a convex combination of its endpoint
loads; the actual equilibrium load lies between the two walls.
Therefore the actual fold loads satisfy
\begin{equation}\label{eq:certified-trapping-width}
 \frac{46219}{100000}<K_+<\frac{46242}{100000},\qquad
 \frac{45897}{100000}<K_-<\frac{45923}{100000},\qquad
 K_+-K_->\frac{37}{12500}>0.
\end{equation}
The final comparison is an exact integer-fraction comparison. In
particular it encloses the full validated load strips, rather than
using the nominal load at an approximate fold as an error estimate.

\section{The global finite certificate}\label{sec:global}
The global part uses thirty-two exact dyadic coefficient arrays, indexed by $0\leq i\leq31$, and two further arrays $U_i^-,U_i^+$ at each index. Their filenames and SHA256 digests are specified in \texttt{global\_replay\_manifest.json}. An array specifies its values on the finite core and the explicit field $Kh+a\eta+K^3\chi$ outside it. Thus it specifies an infinite field, rather than a finite-domain solution. At index $i$ the two wall loads are $K_i\pm1/2000$. The canonical opening $c_i$ is the rational $j/200$ whenever the exact dyadic metadata differ from that rational by less than $10^{-12}$, and is otherwise that exact dyadic metadata. The verifier checks this rule with rational arithmetic and checks strict ordering. In particular,
\[
 c_0=\tfrac14,\quad c_4=\tfrac{29}{100},\quad c_{10}=\tfrac8{25},\quad
 c_{18}=\tfrac25,\quad c_{25}=\tfrac{11}{25},\quad c_{31}=\tfrac{97}{200}.
\]
The complete list of thirty-two rationals is stored in the master certificate. Each physical wall field is changed at the two tip sites by the compact odd lift that makes $\ell(U_i^\pm)=c_i$ exactly. The largest required change is exactly $11/56294995342131200$. Its energy norm is at most $\sqrt{10}$ times its amplitude. The previously proved derivative bounds show that all effects of this change are covered by the explicit $10^{-8}$ allowances in the master. Each auxiliary field $q_i$ is divided by its exact dyadic tip value, so that $\ell(q_i)=1$ exactly.

For $0\leq i<31$ and $0\leq t,s\leq1$, define the physical bilinear seed and auxiliary field
\[
 U(t,s)=(1-t)((1-s)U_i^-+sU_i^+)+t((1-s)U_{i+1}^-+sU_{i+1}^+),
 \qquad q_t=(1-t)q_i+tq_{i+1}.
\]
Its opening is $(1-t)c_i+tc_{i+1}$, independent of $s$, and its load is $(1-t)K_i+tK_{i+1}+(2s-1)/2000$. All coefficients, including the coefficient of $\chi$, are interpolated physically: the latter is the corresponding interpolation of the corner cubes $K^3$, not the cube of the interpolated load.

The exact coefficient checks give $A_V\geq\mu_i I$ on $\XX$ throughout the radius-$1/100$ tube around the whole physical convex hull of the four corners. This is a full-space statement, not only an odd-sector calculation. The constants used by the master are
\[
\begin{array}{c|cccccccc}
 i&0\!:\!1&2\!:\!3&4\!:\!9&10\!:\!17&18\!:\!21&22\!:\!24&25\!:\!26&27\!:\!30\\ \hline
 \mu_i&1/5&1/4&7/20&1/4&3/10&29/100&1/4&1/5.
\end{array}
\]
The checks use exact rational coefficient enclosures and exact integer positive-definiteness certificates on finite Neumann patches. All omitted positive bonds are retained through their proved lower coefficient bound; the negative omitted bonds are bounded by the explicit Gaussian trace tail. Thus no finite Dirichlet eigenvalue is substituted for an infinite coercivity constant.

Here are the master inequalities, which also specify precisely the role of the computed bounds. Write $e_j^\sigma$ for the certified norm of $F(U_j^\sigma)|_{\XX}$, and define
\[
 r_i=\max_{j\in\{i,i+1\}}\bigl\{\text{certified upper bound for }
                 \|A_{U_j}q_j|_{\XX}\|\bigr\},
\]
after normalization. The strip verifier supplies nonnegative constants\[b_c,b_s,b_{qc},b_{qs},b_{qw},j_c,d^2,S_c,S_s,C_1,C_2\]such that
\begin{align*}
 \|F(U(t,s))|_{\XX}\|&\leq \max_{j,\sigma}e_j^\sigma+b_c+b_s=:e,\\
 \|A_{U(t,s)}q_t|_{\XX}\|&\leq r_i+b_{qw}+b_{qc}+b_{qs}=:r,\\
 \|q_{i+1}-q_i\|^2&\leq d^2,\\
 |(A_{U+w}-A_U)(q_t,z)|&\leq C_1\|w\|\|z\|\quad(w,z\in\XX),\\
 |(A_{U+w}-A_U)(q_t,q_t)|&\leq C_2\|w\|.
\end{align*}
All these estimates are uniform on the indicated strip and tube. The symbols include the outward $10^{-8}$ safety additions exactly as implemented. The retained-edge force interpolation bound is obtained from
\[
 \big|\phi'(a+t\delta)-(1-t)\phi'(a)-t\phi'(a+\delta)\big|
 \leq \tfrac18\sup|\phi'''|\,\delta^2.
\]
The analogous differentiated identity gives $b_{qc},b_{qs}$; $b_{qw}$ transfers the base auxiliary residual to the load-wall fields. In the exterior the actual decay $|\phi'''(DU)|=O(r^{-1/2})$ is retained. The midpoint-shell estimate
\[
 \sum_{\text{omitted intact bonds}}r_b^{-p}
 \leq \frac{16}{p-2}(L-1)^{2-p}\qquad(p>2)
\]
with $L=160$, together with the explicit Gaussian remainder below $10^{-70}$, bounds every omitted contribution. The scalar force tail is of order $(L-1)^{-1}$; no faster decay is assumed.

Set $\rho=e/\mu_i$. The master verifies $\rho<9/1000$ in every cell, so the constrained existence lemma gives a unique correction $w(t,s)\in\XX$ in the certified tube with $F(U+w)|_{\XX}=0$. On either load wall put
\[
 \rho_\sigma=\frac{\max(e_i^\sigma,e_{i+1}^\sigma)+b_c}{\mu_i},
 \qquad r_\sigma=r_i+b_{qw}+b_{qc}.
\]
If $f_j^\sigma$ is the certified interval for $F(U_j^\sigma)[q_j]$, then the actual constrained scalar on that wall lies in the convex hull of $f_i^\sigma,f_{i+1}^\sigma$, enlarged by
\[
 j_c+\frac{e_i^\sigma+e_{i+1}^\sigma}{4}\,d
       +r_\sigma\rho_\sigma+\frac{C_1}{2}\rho_\sigma^2.
\]
The term containing $d$ is essential: it accounts for the change from the endpoint test fields to $q_t$. Both wall signs are strict in all cells. The strict load comparison from the analytic part therefore gives a unique smooth equilibrium load for every opening. Adjacent strips have identical physical wall fields at their common opening, so their constrained solutions and load roots agree there and their smooth charts glue.

For completeness, the stability computation uses the true infinite-dimensional Schur complement. Let $Q_-$ and $Q_+$ be the minimum and maximum of the certified corner intervals $A_{U_j^\sigma}(q_j,q_j)$, and let $S=S_c+S_s$ be the certified interpolation error. The quadratic change of the auxiliary field gives the one-sided bounds
\[
 Q_--S-\tfrac14d^2\leq A_U(q_t,q_t)\leq Q_++S.
\]
Consequently the true Schur complement belongs to
\[
 \left[Q_--S-\tfrac14d^2-C_2\rho-
           \frac{(r+C_1\rho)^2}{\mu_i},
       \quad Q_++S+C_2\rho\right].
\]
This is obtained by minimizing over $\XX$; it is not a Rayleigh quotient substituted for the Schur complement. The full-space coercivity on $\XX$ then gives the full Morse index.

The following deliberately outward decimal summaries are consequences of the exact master assertions; the complete thirty-one-row table is included in the archive:
\[
\begin{array}{c|c}
 \text{quantity}&\text{certified bound}\\ \hline
 \max\rho&<0.005668\\
 \text{minimum magnitude of either scalar wall sign}&>0.0010907\\
 \text{stable-region Schur complement}&>0.02138\\
 \text{index-one-region Schur complement}&<-0.03738\\
 \max\rho\text{ on }[0.29,0.32]&<0.002073\\
 \max\rho\text{ on }[0.40,0.44]&<0.003552.
\end{array}
\]
Here the stable regions are $[1/4,29/100]$ and $[11/25,97/200]$, and the index-one region is $[8/25,2/5]$. The two remaining opening intervals are handled by the local-fold certificate. Identification of the local graphs with this global graph uses the separately certified common convex coefficient classes containing both sets of seeds and their radius-$1/100$ tubes: their coercivity constants are $7/20$ on $[29/100,8/25]$ and $29/100$ on $[2/5,11/25]$. The local correction radii are below $0.003041$. Fixed-opening uniqueness in these classes identifies the equilibria even before their loads are known to coincide. No convex class encompassing the whole branch is assumed.

\subsection{Exact one-cell closure}
Let $T$ denote physical translation by one lattice cell in the propagation direction. The terminal convex class is generated by the ten fields
\[
 TU_0^-,\ TU_0^+,\ U_j^-,\ U_j^+\quad(28\leq j\leq31).
\]
For each bond its permitted strain interval is the convex hull of these ten strains enlarged by $\frac1{50}\tau_b$, where $\tau_b=1$ for a reference intact bond and $\tau_{c_n}=\sqrt{2n+3}$ for the omitted seam bond at $x=-n-\frac12$. These are unrestricted energy-space trace bounds: a translated correction need not vanish at the terminal pin. The translated fields are evaluated by reindexing the actual source arrays, including their source core and exterior; they are not replaced by fields with a guessed translated phase.

The exact terminal check has a seventy-dimensional Neumann matrix with all seventy Bareiss pivots positive. Its patch constant is $4/25$, and the omitted negative tail is less than $3.332\cdot10^{-9}<1/100$. The resulting full constrained coercivity is therefore at least $3/20$ throughout the stated convex class.

At opening $c_0$, the constrained correction satisfies
\[
 \|w_0\|<0.001987127.
\]
Since $\ell(w_0)=0$, the sharper translation identities give
\[
 \|Tw_0\|\leq\sqrt3\|w_0\|,
 \qquad |\ell(Tw_0)|\leq\|w_0\|/\sqrt2.
\]
The master, using the exact source-wall values and outward arithmetic, proves
\[
 \|Tw_0\|+10^{-8}<0.003441814<1/50,
 \qquad
 \ell(Tu_0)\in[0.4720462,\,0.4767331]
 \subset(0.47,0.485).
\]
The translated seed is exactly a convex combination of the two translated source-wall fields. Thus these bounds establish its actual membership in the terminal class, without omitting a variable-load predictor correction. The global graph has an equilibrium at this translated opening; its seed and correction also belong to the same class. Fixed-opening uniqueness identifies it exactly with $Tu_0$, including equality of loads. Restricting the global graph to the interval between $c_0$ and this translated opening yields the required one-cell branch. It contains both fold zones and ends in the final stable region. Local uniqueness makes the translated endpoint charts agree as well.

\subsection{Verification provenance}
All inputs, sources, proof notes and recorded inequalities used above are SHA256-bound in \texttt{global\_replay\_manifest.json}. The command \texttt{replay\_global.py combine} checks these bindings and recomputes the thirty-one master inequalities; it consumes the recorded finite certificates. The distinct command \texttt{replay\_global.py replay --workdir NEW} copies only the named immutable files to a fresh directory and recomputes all 191 interval verifier invocations, all 34 global coefficient-hull checks, and the terminal exact check before running the master. The seven expanded local coefficient checks and the local-fold inequalities belong to the separate local replay. Absolute historical paths in descriptive metadata are mapped to the extracted relative identities and are never opened. All actual accesses are relative to the extracted root. A clean relocation checked all 348 named file digests and reproduced every global master assertion. The proof uses the explicit analytic tails and inequalities above in addition to these finite checks; finite truncation alone is not the conclusion.

\section{Completion of the proof}
\begin{proof}[Proof of Theorem~\ref{thm:main}]
The global certificate gives a constrained solution for every point
of the thirty-one continuous opening--load cells, with correction
norm less than $0.005668<0.009<0.01$. The full constrained Hessian
bounds and the projected-residual proposition therefore apply.
The reduced force has opposite strict signs on the two load walls,
and the positive bulk stiffnesses give $R_K<0$. Thus there is a
unique smooth equilibrium load $K(c)$ for each
$c\in[1/4,97/200]$. Neighboring cells have exactly the same canonical
opening and physical wall fields at their shared boundaries, so
the local graphs agree there. Oddness follows from reflection and
uniqueness, while constrained existence and the stability bounds
hold in the full space.

The Schur scalar is positive on $[1/4,29/100]$ and
$[11/25,97/200]$, and negative on $[8/25,2/5]$.
The four narrow local patches establish a simple fold in each of
the remaining zones. The seven expanded patches have exact rational
adjacency and a fixed nonzero cubic sign throughout each zone.
The common convex coefficient classes identify their equilibrium
graphs with the global graph. At every Schur zero the derivative
along the graph equals that cubic, since $K'=0$ there. The
same-orientation crossing lemma consequently excludes any additional
zero in either zone. There are exactly two folds, and the
codimension-one decomposition gives the asserted full Morse indices,
one-dimensional kernels, and coercive complements. In the narrow
fold neighborhoods the complement constant is at least $1/4$.
The load-sign and cubic certificates give both nonzero fold
pairings and the maximum/minimum orientations.

The exact comparisons of all narrow-strip corner loads give the
two displayed threshold enclosures and their strict rational gap.
These comparisons contain the entire validated strips and include
the nonlinear correction and every infinite-tail error through the
preceding certification argument.

Let $u_0$ be the initial exact equilibrium at opening $1/4$.
The terminal certificate puts its exact translate and the global
graph state at the translated opening in one common convex class,
with constrained coercivity $3/20$. The translated opening belongs
to $[0.4720462,0.4767331]\subset(0.47,0.485)$, beyond both fold
zones. Common-opening uniqueness identifies these two states,
including their loads. Restricting the graph to this endpoint gives
the required one-cell segment.

At a regular point, Schur nonvanishing makes the full Hessian
invertible. At a fold, the coercive constrained restriction and
$R_K\ne0$ make the bordered opening--equilibrium derivative
invertible. These are full-space implicit-function statements and
prove the asserted local uniqueness of the arc. Its opening is a
regular coordinate, so the arc itself is regular even at a load
fold. Both endpoints are stable. The translation-join proposition
now supplies a smooth regular parameter and extends the arc to
all real parameter values, with period $T$ modulo one-cell
translation. Hessian congruence and the explicit fold-pairing
identities preserve the two folds and stability sequence in every
period. This proves all claims.
\end{proof}

\appendix
\section{The accompanying certificate archive}
\label{app:archive}

The original verification archive is distributed in two binary parts.
The archive has 606 saved files plus a content manifest and an
assembly note, including the exact binary candidate arrays, all three
verification manifests, the decisive verification sources, recorded
results, and the detailed mathematical derivations. It contains no
download step and no dependency on access to the original workspace.
Its SHA256 digest is
\begin{Verbatim}[fontsize=\small]
7b686d6a8cc084c4d382808733ff00906c470388307b24518c31e90f413ca896
\end{Verbatim}

The candidate data are exact finite definitions of infinite fields;
the tails are the explicit functions proved above. The output JSON
files record interval bounds and exact-matrix certificates, rather
than ordinary floating-point equilibrium approximations. The
arithmetic conventions and analytic bounds in this document explain
why the finite acceptance inequalities imply the infinite-lattice
claims.

Place both files in the same directory and run the following commands
from that directory. The concatenated archive must match the digest above
before extraction. These steps do not execute a verifier.
\begin{Verbatim}[fontsize=\footnotesize]
cat certificate-part01.bin certificate-part02.bin > certificate.zip
python3 - <<'PYTHON'
import hashlib, pathlib
p = pathlib.Path("certificate.zip")
expected = "7b686d6a8cc084c4d382808733ff00906c470388307b24518c31e90f413ca896"
if hashlib.sha256(p.read_bytes()).hexdigest() != expected:
    raise SystemExit("Certificate digest mismatch")
import zipfile
with zipfile.ZipFile(p) as archive:
    archive.extractall("certificate")
PYTHON
\end{Verbatim}

The three saved manifests are
\begin{itemize}
\item \path{scratch/global_translation/global_replay_manifest.json};
\item \path{scratch/variational_order/local_verification_manifest.json};
\item \path{scratch/spectral_exterior/coercivity_manifest.json}.
\end{itemize}
The global manifest lists 191 interval jobs and the selected exact
coefficient checks. The local manifest lists the exact local corner
arrays, all prerequisite certificates, and 55 scalar, interpolation,
cubic, and master commands. The coefficient manifest lists 48 exact
checks. Their union and the supporting sources are included here.
\path{CERTIFICATE_CONTENTS.json} gives a digest and byte count for
every archived source file.

After extraction, the following commands inspect saved-file integrity
without recomputing mathematical bounds:
\begin{Verbatim}[fontsize=\footnotesize]
cd certificate
python3 scratch/global_translation/replay_global.py integrity
python3 scratch/variational_order/replay_local_certificates.py \
  --check-only --check-reference-results
\end{Verbatim}
For a mathematical replay, the saved global driver distinguishes
\verb|combine| (the final inequalities using recorded finite
certificates) from \verb|replay --workdir NEW_DIRECTORY| (fresh
finite interval and exact-rational checks followed by combination).
The local driver requires its force, tangent, and coefficient
prerequisites. Their exact inputs, outputs, programs, and metadata
are archived; the argument conventions are given in
\path{ASSEMBLY_README.txt}. Running
\path{scratch/spectral_exterior/replay_coercivity.py} without
arguments replays its full saved list, and the local driver without
\verb|--check-only| replays its 55 commands. These are optional
reproduction instructions, not a requirement to solve new equilibrium
problems or obtain additional data.

The recorded execution used Python 3.12.4, NumPy 1.26.4, and mpmath
1.3.0. Basic vector interval operations require IEEE-754 binary64,
round-to-nearest, and gradual underflow. The explicit outward
arithmetic and elementary-function enclosures are described in the
proof and verification source. The coefficient positivity tests use
exact integer and rational arithmetic. Floating-point factorizations
only propose witnesses and are not acceptance criteria.

\subsection{Recorded verification and assembly provenance}
The saved final audit covers all 31 global cells, four narrow fold
patches, seven expanded patches, common-class identifications, the
strict threshold gap, and exact translation closure. It includes the
necessary $\sqrt{3/2}$ correction when a projected crack stress bound
is paired with a nonzero-opening lifting. That correction is present
in both saved local master results and in the proof above.

The final assembly reuses these saved results. It does not claim a
new run of the mathematical verification. The global saved relocation
receipt records a successful independent-directory integrity and
master-combination check; completion of further relocated base-check
runs interrupted by the workspace disconnect is not asserted.
The supplied exact inputs and checker programs preserve the finite
evidence supporting the statements independently of that interruption.

Historical component files are retained in the archive for detail and
provenance. In particular, the analytic-only component's original
conditional introduction refers to that component in isolation; all
its hypotheses are supplied by the integrated certificate sections
here. An old residual-method note mentions an exploratory lifting
whose pin was not exactly one; it is not an authoritative final input.
The final exact normalization is the one specified in this manuscript
and the frozen/local input manifests.

\section{Preserved supplementary derivations}
The following supplementary calculations are the saved mathematical
derivations underlying the compact certificate sections. Their plain
text formula notation is retained exactly, with line wrapping only.
In these notes, \texttt{psi} denotes the harmonic tail called $\eta$
in the main text; \texttt{rho} may denote a radial lower bound rather
than a nonlinear correction radius. An upper-case candidate field is
always the exact finite-core/explicit-tail field, not an asserted
equilibrium. Historical numerical examples illustrate an individual
check; the authoritative final results are the manifests and tables
used in the main proof.

\subsection{Detailed residual and stream-tail derivation}
\path{scratch/exact_structure/farfield_and_fold_audit.md}

\begin{Verbatim}[fontsize=\scriptsize]
# Independent audit: corrected far field and augmented fold validation

## A. Corrector and stream: exact identities

Put `a=9/32`, `b=3/64`, `h=r^(1/2) sin(theta/2)`, and

`chi=r^(-1/2)[a log(r) sin(theta/2)+b sin(5theta/2)]`.

Direct differentiation gives

`h_x=-(1/2)r^(-1/2)sin(theta/2)`,
`h_y=(1/2)r^(-1/2)cos(theta/2)`,

and

`3 div(h_x^3,h_y^3)=-(9/32)r^(-5/2)[sin(theta/2)+sin(5theta/2)]`.

For `r^p log(r) sin(m theta)`, the Laplacian is

`r^(p-2)[(p^2-m^2)log(r)+2p]sin(m theta)`.

For `r^p sin(m theta)`, remove the logarithm and `2p` term. These formulas verify exactly that 
`Delta chi=3 div(h_x^3,h_y^3)`.

The stream functions

`psi0=r^(1/2)cos(theta/2)`,
`psi3=-r^(-1/2)[a log(r)cos(theta/2)+b cos(5theta/2)]`

satisfy

`(-psi0_y,psi0_x)=grad h`,
`(-psi3_y,psi3_x)=grad chi-3(h_x^3,h_y^3)`.

Both streams vanish on the negative real axis. Therefore both fluxes have zero slit-normal 
component.

## B. Exact discrete flux and global extension

Use dual lattice `Z^2` for the shifted primal lattice. For an oriented horizontal primal edge, 
take the lower dual endpoint stream value minus the upper one. For an oriented vertical edge, 
take the right dual endpoint value minus the left one. The resulting flux has exactly zero 
discrete divergence at every primal vertex, by cancellation of the four dual values.

Any finite-core extension of the stream, including setting its value at the origin to zero, 
preserves this identity. Every edge touching the modified core must be included in the explicit 
finite residual. Cutting off the flux vector itself, rather than cutting off its stream, would 
generally destroy divergence cancellation.

Taking zero stream values at all nonpositive-axis dual vertices makes the flux across every 
removed crack-space bond zero. These bonds remain present in the physical interaction and their 
actual forces must be bounded separately.

## C. Explicit intact-edge residual bound

Let an intact edge midpoint have radius `r>=R>=2`; put `rho=r-1/2` and `L=1+log(r+1/2)`. Write 
`H=D_e h`, `C=D_e chi`, and let `p0,p3` be the exact discrete fluxes from the streams.

Then

`|H-p0| <= (1/32)rho^(-5/2)`,
`|C| <= (51/128)L rho^(-3/2)`,
`|C-p3-3H^3| <= (663/2048)L rho^(-7/2)`.

Proof details: midpoint quadrature error on a unit segment is bounded by one twenty-fourth of 
the supremum of the integrand's second derivative. All third Cartesian derivatives of `h` have 
magnitude at most `(3/8)rho^(-5/2)`. Using

`chi=-a log(r) Im(z^(-1/2))-b Im(z^(-3/2) conjugate(z))`,

all first Cartesian derivatives of `chi` are bounded by

`[(18log(r)+51)/128] r^(-3/2)`,

and all third Cartesian derivatives by

`[(270log(r)+1413)/512] r^(-7/2)`.

The latter follows from the product rule: for the logarithmic term, the coefficients are 
`(15/8)log(r)+23/4`; for the second term the coefficient is `195/8`. Further,

`|partial_tangent^2(h_e^3)| <= (15/32)rho^(-7/2)`.

Combining the two midpoint errors gives `1413 L/6144+15/256`; replacing `3h_e(midpoint)^3` by 
`3H^3` costs at most `9/256`, all times `rho^(-7/2)`. Their sum is bounded by `(663/2048)L 
rho^(-7/2)`.

For `t=KH+K^3C`, set

`A0=|K|/(2sqrt(rho))`,
`B0=|K|^3(51/128)L rho^(-3/2)`,
`T=A0+B0`.

Since for every real `t`,

`|t exp(-3t^2)-t+3t^3| <= (9/2)|t|^5`,

one obtains the completely explicit estimate

`|f(t)-Kp0-K^3p3| <= |K|rho^(-5/2)/32 + |K|^3(663/2048)Lrho^(-7/2) + 9B0 T^2 + (9/2)T^5`.

Thus the desired improved decay is genuinely justified by a quantitative inequality, not only a 
formal expansion.

For a convenient summable majorant, define

`eta=2 K^2(51/128)[1+log(R+1/2)]/(R-1/2)`,
`A=|K|(1+eta)/2`,
`C0=|K|/32+(9/2)A^5`,
`C1=|K|^3[(663/2048)/(R-1/2)+9(51/128)A^2]`.

The edge residual is at most `[C0+C1 L]rho^(-5/2)`. For integer `N>=R>=2`, the number of 
oriented-positive edge midpoints with `n<=r<n+1` is at most `57n`. This follows by covering the 
centers with disjoint unit squares for each of the two edge orientations and enclosing them in 
the annulus of radii `n-1,n+2`.

Consequently, the squared intact-edge tail is bounded by

`57(4/3)^5 sum_(n>=N) [C0+C1(2+log n)]^2 n^(-4)`.

Set `D=C0+2C1` and `V=D+C1 log N`. Since the summand is decreasing, the sum is no larger than

`V^2/N^4 + [V^2+(2C1/3)V+2C1^2/9]/(3N^3)`.

This is an explicit `R^(-3)log^2(R)` bound for the squared residual norm.

## D. Exponentially small physical crack-tail forces

For a cut bond with midpoint `(-s,0)`, where `s=n+1/2>=3/2`, the upper endpoint has `theta in 
(4pi/5,pi)` and `log r>0`. Thus `chi>0` there, and the corrected predictor jump is at least `2K 
sqrt(s)` for `K>0`.

Suppose `K>=k0>0`, and an energy-norm correction ball of radius `epsilon` is centered on a 
predictor-only exterior. The intact detour between the cut-bond endpoints has length `2s+2`. 
Therefore the perturbation of the jump is at most `epsilon sqrt(2s+2)`.

For `s>=s_N`, put

`beta=2k0-epsilon sqrt(2+2/s_N)`.

If `beta>0` and `beta sqrt(s_N)>=1/sqrt(6)`, then monotonicity of `f` on its decreasing positive
 branch gives

`0<f(jump)<= beta sqrt(s) exp(-3beta^2 s)`.

The cut-bond dual norm is bounded by summing this quantity times `sqrt(2s+2)`. Since 
`sqrt(s(2s+2))<=sqrt(2+2/s_N)s`, the tail is bounded by an explicit multiple of

`sum_(n>=N) (n+1/2) exp(-a(n+1/2))`, where `a=3beta^2`.

This sum equals

`exp(-a(N+1/2))[(N+1/2)/(1-exp(-a))+exp(-a)/(1-exp(-a))^2]`.

## E. Candidate and cutoff cautions

The improved tail cannot simply be grafted onto an old finite solution with uncorrected boundary
 data while keeping its interior displacement fixed. The mismatch is of size 
`chi(R)~log(R)/sqrt(R)` across a one-ring interface. There are order `R` such bonds, so that 
sharp graft can create an energy-norm defect of order `log(R)`. One must re-solve using the 
corrected boundary data, use a controlled transition, or explicitly compute and certify every 
transition residual.

The force residual improvement also does not improve the leading Hessian perturbation, which 
remains order `K^2/R` in the far intact lattice.

Finally, the null-vector residual requires an independent estimate. A possible leading 
kernel-mode tail `q~c r^(-1/2)sin(theta/2)` has omitted energy norm of order `R^(-1/2)`. 
Improved equilibrium asymptotics alone do not improve this eigenvector truncation error.

## F. Actual augmented-fold inverse estimate

Work in the Hilbert energy space and identify its dual using the reference Riesz map. Let

`G(v,K,q)=(F(v,K), H(v,K)q, (||q||^2-1)/2)`.

Assume first that `||q||=1`, `Hq=0`, and that `H` restricted to `q^perp` is invertible with 
inverse norm at most `1/gamma`. Put `P=I-q tensor q`,

`f=F_K`, `mu=<q,f>`,
`B[a,b]=F_vv[a,b]`, `beta=<q,B[q,q]>`,
`h=F_vK q`.

For input `(y,z,c)`, write the solution of `DG(a,k,b)=(y,z,c)` as `a=alpha q+aC`, `b=cq+bC`. 
Exact block elimination gives

`k=<q,y>/mu`,
`aC=H_C^(-1)(Py-Pf k)`,
`alpha=[<q,z>-<P B[q,q],aC>-<q,h>k]/beta`,
`bC=H_C^(-1)(Pz-P B[aC,q]-alpha P B[q,q]-Ph k)`.

Thus merely knowing `mu,beta,gamma` does not bound the inverse. Define

`fC=||Pf||`, `bC0=||P B[q,q]||`,
`BCC=||P B[P(.),q]||`, `hC=||Ph||`, `hq=|<q,h>|`.

For the maximum norm on the three product factors, set

`U=(1+fC/|mu|)/gamma`,
`V=(1+bC0 U+hq/|mu|)/|beta|`,
`Z=(1+BCC U+bC0 V+hC/|mu|)/gamma`.

Then a valid inverse bound is

`M0=max(1/|mu|,sqrt(U^2+V^2),sqrt(1+Z^2))`.

These formulas retain useful cancellations and are substantially better than replacing every 
cubic quantity by the same global derivative bound.

For an approximate normalized vector, put `eta_q=||Hq||` and define `H0=P H P`. Since 
`||H-H0||<=2eta_q`, the corresponding idealized augmented inverse `A0` is a true local inverse 
perturbation if `2M0 eta_q<1`, with

`||(DG)^(-1)|| <= M0/(1-2M0 eta_q)`.

Alternatively, use `A0` directly in a contraction certificate and account for the derivative 
defect `delta<=2M0 eta_q`.

## G. Defect threshold and genuinely missing numerical quantities

Let `epsilon_G=||G(z0)||`, let `||A0||<=M`, let `||I-A0 DG(z0)||<=delta<1`, and let `DG` be 
Lipschitz with constant `L` on the proposed validation ball. The Newton-like map is certified by
 a radius `r` satisfying

`M epsilon_G + (delta-1)r + (M L/2)r^2 < 0`,
`delta+MLr<1`.

A necessary discriminant condition for these particular estimates is

`epsilon_G < (1-delta)^2/(2 M^2 L)`.

The reported force-tail values `0.000594`, `0.000218`, and `0.000081` cannot be compared with 
this threshold until all of the following have been certified:

1. Total force residual, including core rounding and all boundary-interface contributions.
2. Total kernel-vector residual `||Hq||` and normalization error.
3. The six block quantities `fC,bC0,BCC,hC,hq,gamma` on the same infinite-space formulation.
4. Rigorous lower bounds for `|mu|` and `|beta|`, including finite-tail and rounding errors.
5. A local Lipschitz bound `L` for the augmented derivative, including physical seam traces and 
parameter derivatives.
6. A radius for which the space, predictor, derivative bounds, and positive lower loading bound 
remain valid.

For illustration only, if `delta=0` and `L=100`, the three reported defects would require 
`M<2.90`, `M<4.79`, and `M<7.86`, respectively. These are not certificates: `L=100` has not been
 established, and these numbers ignore the potentially dominant kernel-vector residual.

At an approximate fold, changing from affine loading `Kh+v` to corrected loading `Kh+K^3chi+w` 
changes `F_K` by `H(3K^2chi)`. Therefore the projected transversality changes by `<Hq,3K^2chi>`.
 It vanishes at an exact kernel but cannot be dropped at an approximate kernel.

No claim in this audit proves existence of an infinite-lattice fold or the full branch.

## H. A sharper complement bound from codimension-one coercivity

This is an independently proved improvement over a geometric angle bound.

Let `H` be bounded self-adjoint on the energy Hilbert space. Suppose a closed codimension-one 
subspace `Y` satisfies

`<Hw,w> >= delta ||w||^2` for every `w in Y`, with `delta>0`.

Let `||q||=1` and `eta=||Hq||<delta`. Then there is exactly one eigenvalue of `H` in 
`(-infinity,delta)`, it has multiplicity one, and its value `lambda` belongs to `[-eta,eta]`. 
Furthermore,

`<Hw,w> >= [delta-(delta+eta)(eta/delta)^2] ||w||^2`

for every `w` orthogonal to `q`.

Proof. The spectral subspace corresponding to `(-infinity,delta)` has dimension at most one. 
Otherwise its intersection with `Y` contains a nonzero vector, whose quadratic form is strictly 
below `delta` times its squared norm, contradicting the assumed coercivity. The inequality 
`dist(0,spectrum(H))<=||Hq||=eta<delta` implies that this spectral subspace is nonzero. It is 
therefore one-dimensional, with a normalized eigenvector `e` and eigenvalue `lambda in 
[-eta,eta]`.

Write `q=alpha e+r`, where `r` is in the spectral subspace for `[delta,infinity)`. Orthogonality
 of spectral subspaces gives `eta^2=||Hq||^2>=delta^2||r||^2`. For any unit vector `w` 
orthogonal to `q`, its coefficient along `e` is bounded in modulus by 
`sqrt(1-|alpha|^2)=||r||<=eta/delta`. Hence

`<Hw,w> >= lambda |<w,e>|^2+delta(1-|<w,e>|^2) >= delta-(delta+eta)(eta/delta)^2`.

At an exact kernel, this gives complementary coercivity at least `delta` itself. Thus certified 
codimension-one constants such as `0.748` and `0.419` could be substantially more useful than 
coarse complementary bounds `0.061` and `0.120`. The full infinite-lattice residual `eta=||Hq||`
 remains indispensable.

## I. Audit of a finite Neumann-patch spectral certificate

The constants `0.748` and `0.419` mentioned above were subsequently clarified to be losses in a 
different reference-form comparison, not coercivity constants in the energy norm. They cannot be
 used as `delta` in Section H. Only a proved lower bound in the actual energy inner product is 
admissible there.

A separate finite-patch route is rigorous under the following exact hypotheses. Let `C` be a 
finite vertex patch containing both endpoints of the constraint bond `p`. Let `E_C` contain 
every physical edge with both endpoints in `C`; distinguish intact reference edges from seam 
edges. Define Neumann forms by summing only these internal edges. Suppose

`H_C[w,w] >= mu L_C[w,w]` whenever `D_p w=0`.

Assume every omitted intact edge, including every edge crossing the patch boundary, has 
coefficient at least `mu`. Suppose the negative omitted seam terms satisfy

`sum_(omitted seam e) max(0,-c_e)|D_e w|^2 <= tau L0[w,w]`.

Then for every global finite-energy `w` satisfying `D_p w=0`,

`H[w,w] >= (mu-tau)L0[w,w]`.

Proof. Apply the finite form inequality to the restriction of `w` to `C`, bound every omitted 
intact contribution below by `mu` times its reference contribution, retain nonnegative omitted 
seam contributions, and bound the remaining negative seam sum by `tau L0`.

For validation, replace each finite coefficient by an exact rational lower bound. Identify the 
two endpoints of `p` to enforce the constraint and fix one vertex to remove the constant. A 
verified positive-semidefinite LDL factorization of the resulting matrix `H_C-mu L_C` proves the
 finite inequality. An ordinary floating-point LDL computation is exploratory only. A 
Dirichlet-pinned finite matrix cannot replace this Neumann matrix: its boundary springs add 
positive energy unavailable in the stated lower-bound argument.

If the negative tail starts at cut coordinate `s_N` and its opening is at least `beta sqrt(s)` 
with `beta sqrt(s_N)>=1/sqrt(2)`, then

`max(0,-phi''(jump)) <= (6beta^2 s-1)exp(-3beta^2 s)`.

Combining this with the detour trace bound gives an explicit convergent quadratic 
geometric-series bound for `tau`. On the constraint space `D_p w=0`, one can shorten the trace 
path by terminating both paths at the endpoints of `p`; the jump on `p` contributes zero.

A failed finite Neumann PSD certificate does not refute global coercivity: dropping positive 
exterior edges weakens the form. A successful certificate with all stated boundary and tail 
estimates is a valid global bound.
\end{Verbatim}

\subsection{Detailed interpolation-tail derivation}
\path{scratch/exact_structure/interpolation_certificate_method.md}

\begin{Verbatim}[fontsize=\scriptsize]
# Uniform interpolation defect certificates

Let two explicit infinite candidate fields be `U0,U1`, and two explicit finite-energy lifting 
fields be `Q0,Q1`. Set `Ut=(1-t)U0+tU1` and `Qt=(1-t)Q0+tQ1`, for `0<=t<=1`. Interpolate the 
exact divergence-free endpoint streams by the same weights. Their discrete curls remain 
divergence-free. Their tip-annihilator coefficients also interpolate, which is harmless on `ker 
D_tip`.

On one edge put `dj=D Uj`, `ej=D Qj`, `Delta d=d1-d0`, and `Delta e=e1-e0`. The scalar 
interpolation remainder gives

`|phi'(dt)-(1-t)phi'(d0)-t phi'(d1)| <= M3 |Delta d|^2/8`,

where `M3` bounds `|phi'''|` on the strain interval. Differentiating `phi''(dt)et` twice with 
respect to `t` gives

`phi''''(dt)(Delta d)^2 et+2phi'''(dt)Delta d Delta e`.

Thus the tangent traction interpolation defect is bounded by

`[M4 |Delta d|^2 max(|e0|,|e1|)+2M3|Delta d||Delta e|]/8`.

The interval program uses the exact derivative identities

`phi'''(d)=18d(2d^2-1)exp(-3d^2)`,
`phi''''(d)=18(-12d^4+12d^2-1)exp(-3d^2)`.

Every finite edge through `L=256` is enclosed by the vector interval engine. For the full-space 
bound all intact edges are included. For the projected bound the tip bond is omitted because it 
annihilates `ker D_tip`. Physical negative seam bonds are bounded using their detour trace 
norms, rather than being deleted.

For the intact exterior let `rho=r-1/2` and `N=L`. For endpoint parameters `Kj,aj,bj` 
(`bj=aprime_j`) define

`Aj=Kj/2+[Kj^3(51/128)(1+log(N+1/2))+|aj|/2]/(N-1/2)`,
`A=max(A0,A1)`,
`B=|K1-K0|/2+[|K1^3-K0^3|(51/128)(1+log(N+1/2))+|a1-a0|/2]/(N-1/2)`,
`Q=max(|b0|,|b1|)/2`, `E=|b1-b0|/2`.

Then `|dt|<=A rho^(-1/2)`, `|Delta d|<=B rho^(-1/2)`, `|et|<=Q rho^(-3/2)`, and `|Delta e|<=E 
rho^(-3/2)`. In particular,

`M3<=C3 rho^(-1/2)`, `C3=18A[1+2A^2/(N-1/2)]`,
`M4<=C4`, `C4=18[1+12A^2/(N-1/2)+12A^4/(N-1/2)^2]`.

Consequently the force interpolation defect is at most `Cf rho^(-3/2)` with `Cf=C3 B^2/8`, and 
the tangent defect is at most `Ct rho^(-5/2)` with `Ct=(C4 B^2 Q+2C3 B E)/8`.

The full-plane shell count of at most `57n` edge midpoints in `n<=r<n+1` gives the squared tail 
bounds

`57(4/3)^3 Cf^2 [N^(-2)+N^(-1)]`,
`57(4/3)^5 Ct^2 [N^(-4)+1/(3N^3)]`,

respectively. Notice that using a constant global bound for `M3` in the force estimate would 
fail to give a summable exterior majorant when `K0!=K1`. The decay of the third derivative is 
essential.

On the omitted physical seam use twice the maximum endpoint-family force/tangent tail, with 
`Kmin=min(K0,K1)` and `bmax=max(|b0|,|b1|)`. The linear interpolant has opening at least `2Kmin 
sqrt(s)` because both phases are nonnegative. The explicit geometric-series bounds in 
`residual_certificate_method.md` apply, with the extra factor two coming from the difference 
between the interpolated traction and the traction of the interpolated field.

The resulting JSON values are uniform in the entire segment parameter `t`. They must be added to
 appropriate endpoint residual bounds; they are not equilibrium or branch certificates by 
themselves.
\end{Verbatim}

\subsection{Detailed scalar-tail derivation}
\path{scratch/variational_order/scalar_tail_bounds.md}

\begin{Verbatim}[fontsize=\scriptsize]
# Explicit scalar tail estimates for the matched predictor

This is an analytical estimate for the explicit approximate fields, not an assertion that they 
solve the nonlinear equilibrium equations.

Assume 0.45<=K<=0.5, 0<=a<=1/3 and 0<=b<=1.1. Outside the computed core put

    U=K h+K^3 chi+a psi,   q=b psi,
    h=r^(1/2) sin(theta/2), psi=r^(-1/2) sin(theta/2),
    chi=(9/32)r^(-1/2)log(r)sin(theta/2)+(3/64)r^(-1/2)sin(5theta/2).

The discrete point force and the two differentiated point forces are

    f_i=sum_{j~i} phi'(U_i-U_j),
    t_i=sum_{j~i} phi''(U_i-U_j)(q_i-q_j),
    g_i=sum_{j~i} phi''(U_i-U_j)(h_i-h_j).

At all sites of radius r>=512, the following conservative envelopes hold:

    |f_i| <= (1+0.1 log r)r^(-7/2),
    |t_i| <= 2r^(-7/2),
    |g_i| <= 2r^(-5/2),
    |q_i| <= 1.1r^(-1/2).

Exponentially small physical cross-crack terms can be included in these envelopes using the 
strict margins in the estimates below.

## Derivative bounds

For any unit Cartesian direction and k=1,2,3,4, pure kth derivatives of h are bounded by H_k 
r^(1/2-k), with

    (H_1,H_2,H_3,H_4)=(1/2,1/4,3/8,15/16).

For psi the corresponding constants in r^(-1/2-k) are

    (P_0,P_1,P_2,P_3,P_4)=(1,1/2,3/4,15/8,105/16).

For f=r^(-1/2)sin(5theta/2), write f=Im(z conjugate(z)^(-3/2)); the product rule gives 
derivative constants

    (D_0,D_1,D_2,D_3,D_4)=(1,5/2,27/4,195/8,1785/16).

The logarithmic product rule therefore yields

    |partial_e^k chi| <= r^(-1/2-k) C_k(log r),

where

    C_1(l)=(9/64)l+51/128,
    C_2(l)=(27/128)l+225/256,
    C_3(l)=(135/256)l+1413/512,
    C_4(l)=(945/512)l+11691/1024.

For example these follow from

    C_k(l)=(9/32)[P_k l+sum_{j=1}^k binom(k,j)P_(k-j)(j-1)!]+(3/64)D_k.

On every unit stencil centered at radius r>=512, monotonicity of (A log r+B)/r and the preceding
 formulas give the convenient bounds

    |U_e| <= .252 r^(-1/2),
    |U_ee| <= .128 r^(-3/2),
    |U_eee| <= .193 r^(-5/2),
    |U_eeee| <= .483 r^(-7/2),
    |h_e| <= .501 r^(-1/2),
    |h_ee| <= .252 r^(-3/2),
    |h_eeee| <= .945 r^(-7/2),
    |q_e| <= .552 r^(-3/2),
    |q_ee| <= .830 r^(-5/2),
    |q_eeee| <= 7.3 r^(-9/2).

Each decimal is an exact rational upper bound. To verify uniformly, evaluate at the smallest 
stencil radius 511 the ratio to the displayed power of 512. The ratios decrease thereafter. The 
required logarithmic inequality follows from the exact rational inequality 
sum_{k=0}^{30}(6237/1000)^k/k!>511. Squaring the positive derivative bounds leaves purely 
rational comparisons. For example the U kth derivative comparison is

    [(1/2)H_k+((1/8)C_k(6237/1000)+(1/3)P_k)/511]^2
       (512/511)^(2k-1) < (displayed rational bound)^2.

The h and q comparisons similarly use H_k^2(512/511)^(2k-1) and [(11/10)P_k]^2(512/511)^(2k+1). 
All ten comparisons were checked with exact integer-fraction arithmetic; their results are 
recorded in tail_rational_checks.json. No rounded logarithm is needed for these endpoint checks.

## Point-force cancellation

The continuum identity

    Delta chi=9 sum_{e=e1,e2} h_e^2 h_ee

is exact. Both sides equal -(9/32)r^(-5/2)[sin(theta/2)+sin(5theta/2)]. Also Delta h=Delta psi=0
 away from the cut.

For one coordinate let t_+=U(x+e)-U(x), t_-=U(x)-U(x-e), p=U_e(x), d=t_+-t_-, and m=(t_++t_-)/2.
 If M_k bounds |partial_e^k U| on the stencil, then

    |d-U_ee(x)|<=M_4/12, |m-p|<=M_3/6, |d|<=M_2,
    3(t_+^3-t_-^3)=9m^2 d+(3/4)d^3.

Consequently the difference between 3(t_+^3-t_-^3) and 9p^2 U_ee is bounded by

    (3/2)M_3(2M_1+M_3/6)M_2+(3/4)M_1^2 M_4+(3/4)M_2^3.

Writing phi'(t)=t-3t^3+R(t), Taylor's theorem applied to (1-6s)exp(-3s) gives |R'(t)|<=22.5t^4 
for t^2<=1/6. Thus the remaining paired force contributes at most 22.5M_1^4 M_2 in each 
direction.

Set V=U-Kh. At the center,

    |V_e|<=B_1(log r)r^(-3/2),
    |V_ee|<=B_2(log r)r^(-5/2),
    B_1(l)=(9/512)l+665/3072,
    B_2(l)=(27/1024)l+737/2048.

Here 665/3072=.216471354... and 737/2048=.35986328125. For r>=512, B_1/r<.00064 and 
B_2/r<.00103. Expanding U_e^2 U_ee-(Kh_e)^2(Kh_ee), summing directions, and multiplying by9 
bounds the uncancelled continuum cubic term by

    (.66+.05 log r)r^(-7/2).

The linear central-difference error is at most (.483/6)r^(-7/2)=.0805r^(-7/2). The paired cubic 
discretization errors sum to less than .088r^(-9/2). The fifth-order remainder sums to at most

    45(.252)^4(.128)r^(-7/2)<.0233r^(-7/2).

Their total is strictly smaller than (1+.1 log r)r^(-7/2).

## Differentiated-force estimates

For |t|<=1/sqrt(6), |phi''(t)-1|<=9t^2 and |phi'''(t)|<=18|t|. Express the edge correction 
(phi''(DU)-1)Dh at edge midpoints and differentiate in the edge direction. Its derivative is 
bounded by

    18 M_1(U) M_2(U) M_1(h)+9 M_1(U)^2 M_2(h).

Summing two directions and adding the harmonic h discretization error gives |g_i|<2r^(-5/2). 
Replacing h by q gives

    2[18(.252)(.128)(.552)+9(.252)^2(.830)]r^(-7/2)
      +(7.3/6)r^(-9/2) <2r^(-7/2),

which proves the t_i envelope.

At the negative crack face one applies these smooth derivative estimates to the even reflection 
of the upper-branch expressions. The reflected missing neighbor has the same value as the 
center, so this estimates the deleted-stencil part exactly. The full physical seam is then 
restored. Its opening Q is at least .88sqrt(r) and at most2sqrt(r) for r>=512. The extra force 
is at most2sqrt(r)exp(-2r). The degree-five term of exp(2r) gives the bound 
(7.5/r)r^(-7/2)<.015r^(-7/2). The extra tangent force is at most2.2(1+24r)r^(-1/2)exp(-2r); the 
degree-six exponential term bounds it by .003r^(-7/2). The same calculation bounds the extra 
loading derivative by .003r^(-5/2). These fit the strict margins above. There is no deletion of 
a physical interaction in the resulting estimate.

## Summed scalar tails

The full lattice shell max(|x1|,|x2|)=n+1/2 has 8n+4=8s vertices, s=n+1/2. Suppose an upper-half
 site sum includes |x1|<L and 0<x2<L, and use reflection to obtain the full scalar pairing. Set 
A=L-1/2>=511.5. Absolute convergence and comparison with a decreasing integral give

    omitted |<F,q>| <= [4.62+.44 log(sqrt(2)A)]/A^2,
    omitted |<Aq,q>| <= 8.8/A^2,
    omitted |<F_K,q>| <= 17.6/A.

For L=2048 these bounds are less than 2.0e-6, 2.11e-6, and .00860 respectively. They bound the 
physical infinite tails, not just a numerical truncation difference.

Finally |phi'''(DU)(Dq)^3|<=18(.252)(.552)^3 r^(-5) on every intact tail edge. The four directed
 terms over upper sites give at most [16*18*.252*.552^3/3]A^(-3). The physical negative seam 
contributes at most3090exp(-2r) per seam term. Using the degree-eight exponential term and 
summing gives at most69525A^(-7), which is smaller than the spare coefficient between 
16*18*.252*.552^3/3 and4.1 times A^(-3), since A>=511.5. Hence the complete cubic-pairing tail 
is bounded by4.1/A^3, and at L=2048 is less than5e-10.

## Wider parameter box for the whole-cell mesh

The same proof works on 0.45<=K<=0.5, 0<=a<=2/5, 0<=b<=7/5. All four displayed U derivative 
bounds remain valid (the positive rational squared comparisons at radius511 have been rechecked 
with a=2/5). The q derivative bounds may be replaced by .704r^(-3/2), 1.06r^(-5/2), 
and9.3r^(-9/2). The corresponding rational comparisons use (7P_k/5)^2(512/511)^(2k+1). For the 
force cancellation, B_1 and B_2 gain respectively1/30 and1/20 relative to the local box; the 
original force envelope still has sufficient margin, but the following larger common envelopes 
are convenient:

    |f_i| <= (2+.2log r)r^(-7/2),
    |t_i| <= 3r^(-7/2),
    |g_i| <= 3r^(-5/2),
    |q_i| <= 1.4r^(-1/2).

They give the global scalar tails

    omitted |<F,q>| <= [11.76+1.12log(sqrt(2)A)]/A^2,
    omitted |<Aq,q>| <= 16.8/A^2,
    omitted |<F_K,q>| <= 33.6/A,
    omitted cubic pairing <= 12/A^3.

The intact cubic coefficient is16*18*.252*.704^3/3<9, leaving ample margin below12 for the 
physical negative seam contribution. The verifier selects this larger box whenever a candidate 
is outside the local box.
\end{Verbatim}

\subsection{Detailed global-cell inequalities and tails}
\path{scratch/global_translation/cell_strip_certificate_proof.md}

\begin{Verbatim}[fontsize=\scriptsize]
# Interpretation and tail proof for `verify_cell_strip.py`

The script supplies constants for a bilinear family of explicit physical
seeds, not by itself a solution certificate. Its four corners are the
two endpoint openings, each at the two prescribed load walls. The
interpolated physical field is the affine/bilinear combination of the
corner fields. In particular, its coefficient of chi is the same
combination of the corner K-cubes; it is not recomputed as the cube of
the interpolated K.

Normalize each stored q by its exact dyadic tip value, so q(p+)=1
exactly. Let q_t be the affine interpolation of the two base q fields.
The correction space is the odd fixed-opening subspace. Thus every
correction w has D_p w=0, and on c_n its trace norm is at most
sqrt(2n+2), using the two horizontal paths to the two zero pin values.

## Finite calculations

All finite operations use the outward-rounded interval engine already
used by the force certificates. The script encloses phi''' and phi''''
over the coefficient interval containing the four wall fields, the two
base fields, and the radius-.01 fixed-opening correction tube. Exact
critical-point formulas for these derivatives are included when they
fall in the interval. The pin bond is omitted from all dual/state-error
constants, but retained in scalar interpolation along changing opening.

If m3 and m4 are the resulting coefficient bounds, DC bounds the
physical field difference between the two openings on either wall,
DK bounds the difference between the two load walls, Q bounds |Dq_t|,
and DQ=|D(q_1-q_0)|, the named stress bounds are:

* force_c: m3 DC^2/8;
* force_load: m3 DK^2/8;
* Hq_c: m4 DC^2 Q/8 + m3 DC DQ/4;
* Hq_load: m4 DK^2 Q/8;
* Hq_wall_difference: m3 |D(U_wall-U_base)| |Dq_base|;
* C2q: m3 Q^2, converted to a dual norm.

The conversion is the square root of the sum of squared intact-bond
coefficients, plus the sum of each deleted-edge coefficient times
sqrt(2n+2). Full-lattice bulk sums are twice upper-half sums; seam
sums occur once. `Cq` is instead

    max_{B0 except p} m3 Q + sum_{c_n} (2n+2)m3 Q.

It controls both the scalar quadratic correction and the operator
change Hq under a fixed-opening state correction.

The scalar force interpolation contribution is sum(m3 DC^2 Q)/8.
To compare with endpoint scalars F_i[q_i], the master certificate must
also add

    (eta_F,0+eta_F,1) ||q_1-q_0|| / 4,

because q changes. This uses q_1-q_0 in the fixed-opening space.

The Schur interpolation losses are

    schur_c_first = sum(m4 DC^2 Q^2)/8,
    schur_c_cross = sum(m3 DC DQ Q)/2,
    schur_load = sum(m4 DK^2 Q^2)/8.

There is additionally ||q_1-q_0||_H^2/4 in the lower bound. Indeed,
writing Abar=(1-t)A0+tA1 and qt=(1-t)q0+tq1 gives the exact identity

    A_t[qt,qt] - [(1-t)A0[q0,q0]+t A1[q1,q1]]
      = (A_t-Abar)[qt,qt]
        -2t(1-t)(A1-A0)[qt,q1-q0]
        -t(1-t)[t A0+(1-t)A1][q1-q0,q1-q0].

The last term is nonpositive by constrained coercivity. It is bounded
below by -||q1-q0||_H^2/4 because phi''<=1 on every retained bond and
all deleted-seam coefficients are negative throughout the box. The
script checks the latter fact. Thus that term is omitted from the
upper Schur bound, but not the lower one.

## Elementary global derivative bounds

For all real x,

    |phi'''(x)| <= 18|x|,    |phi''''(x)| <= 18.

For the second assertion put s=x^2 and P=-1+12s-12s^2. Since
exp(3s)>=1+3s+(9/2)s^2,

    exp(3s)-P >= 2-9s+(33/2)s^2 > 0

(the quadratic has negative discriminant). The cubic exponential
partial sum also gives

    exp(3s)+P >= (3/2)s(10-5s+3s^2) >= 0.

Hence |P|exp(-3s)<=1 and the bound follows from
phi''''(x)=18P exp(-3s). The floating critical-value sanity check in
the script is not the justification for using the exact tail bound18.

## Exterior power envelopes

All finite sums retain bulk edge midpoints in the rectangle of radius
L=160. Assign omitted edges to a shifted-lattice endpoint; sacrificing
one unit leaves A=159 as a safe shell lower bound.

On this exterior, for every corner and every bilinear combination,

    |DU| <= r^(-1/2),       |Dq_t| <= r^(-3/2),
    |D Delta_c U|, |D Delta_load U|,
    |D(U_wall-U_base)| <= .01 r^(-1/2),
    |D(q1-q0)| <= .3 r^(-3/2).

Here r refers to the assigned endpoint radius, with the one-unit
stencil margin already absorbed. To verify these generous envelopes,
use |grad h|=1/(2sqrt(r)), |grad psi|=1/(2r^(3/2)), and

    |partial_e chi| <= [(9/64)log r+51/128] r^(-3/2).

Every corner has .45<K<.5, chi coefficient at most1/8, and 0<a<.4.
Within one cell the K spread is less than .002, hence the spread of
the corner chi coefficients is less than .0015. Even the crude
phase spread .4 yields an intact-gradient difference below

    [.002*.502 + .0015*.0071 + .4*.505/159] r^(-1/2)
      < .01 r^(-1/2).

The derivative ratios decrease beyond159. The q amplitudes are at
most1.4, giving |Dq|<.708r^(-3/2)<r^(-3/2). The script checks that
the adjacent normalized q-tail amplitudes differ by less than .3;
the derivative factor is below.505, so the last displayed envelope
has ample slack. These bounds also hold for affine combinations of
the coefficients, without imposing the relation b_chi=K^3 inside a
cell.

A radius-.01 correction changes an intact strain by at most rho=.01.
Thus in third-derivative state-error bounds one uses

    m3 <=18(rho+r^(-1/2)).

For omitted bond midpoints, the infinity-radius levels are
L+1/2,L+1,L+3/2,..., with8r bond midpoints per half-step at radius r.
Each bond segment has Euclidean radius at least r-1/2. A decreasing
sum/integral comparison, with one unit of radius sacrificed, gives
for p>2

    sum (segment_radius)^(-p) <=16/(p-2) (L-1)^(2-p).

For example put s=r-1/2 and integrate from L-1/2; the resulting
extra term from r=s+1/2 is absorbed by replacing that lower limit
by L-1 (Bernoulli's inequality gives the same comparison directly).
Thus with A=L-1 the convenient full-edge sums are

    sum r^-3 <=16/A,
    sum r^-4 <=8/A^2,
    sum r^-5 <=16/(3A^3),
    sum r^-6 <=4/A^4.

These follow from the decreasing integral comparison after replacing
the shell lower endpoint by A. All shell counts in the code are
upper bounds; boundary-crossing edges are retained explicitly.

For B=.01, C=.3, these estimates give the tail additions appearing in
the script:

    force_c, force_load <=9B^2/sqrt(A),
    Hq_c <=[(18/8)B^2+(18/4)BC]sqrt(16/(3A^3)),
    Hq_load <=(18/8)B^2 sqrt(16/(3A^3)),
    Hq_wall_difference <=18(rho+1/sqrt(A))B sqrt(8/A^2),
    C2q <=36(rho+1/sqrt(A))/A^2,
    Cq <=18(rho+1/sqrt(A))/A^(3/2).

The scalar force interpolation tail is36B^2/A. In particular it
decays as A^-1, not A^-2. The three Schur tails are respectively
18B^2/A^2,72BC/A^2,18B^2/A^2. The q-difference norm tail uses the
actual interval-enclosed difference d of its harmonic amplitudes,
with bound10d^2(A^-1+A^-2).

## Deleted-seam Gaussian tails

For s=n+1/2>=160.5 all corner predictors have opening at least
.9sqrt(s), since K>=.45 and their chi/psi corrections are positive
there. Subtracting the radius-.01 trace error leaves opening at least
.88sqrt(s). A generous upper bound is2sqrt(s). Also
|Dq_t|<=3/sqrt(s), |D Delta q|<=6/sqrt(s), and either physical field
difference is at most4sqrt(s). Thus

    m3 <=324s^(3/2) exp(-2s),
    m4 <=4338s^2 exp(-2s).

Every omitted weighted term used by the script is consequently at
most10^7(n+1)^3 4^-n, using e>2. The ratio of successive terms is
less than1/2 for n>=160, and

    2*10^7*161^3*10^70 < 4^160.

Therefore every complete deleted-seam remainder is below1e-70.
No physical interaction is removed.

## Master implications and normalization margin

For a cell with four corner projected-force bounds eta_i and a
coercivity certificate mu, put

    eta=max eta_i+force_c+force_load+rounding_margin,
    rho=eta/mu.

Require rho<.009; the coefficient constants were proved on radius.01.
This yields a constrained solution over the entire bilinear strip.
At either wall the exact reduced scalar differs from its seed scalar
by at most

    rq*rho + Cq*rho^2/2,
    rq=max base_eta_Hq+Hq_wall_difference+Hq_c+Hq_load.

Use the scalar interpolation and changing-q terms above to establish
opposite signs uniformly on both walls. For stability, the actual
Schur complement lies between

    min corner_Q - schur_c_first-schur_c_cross-schur_load
      -deltaq_norm_squared/4-C2q*rho-(rq+Cq*rho)^2/mu,
    max corner_Q + schur_c_first+schur_c_cross+schur_load+C2q*rho.

The finite files use exact stored pin values. To define one common
physical-opening family, correct each corner by
(c_i-U_i(p+))*zeta. After explicitly checking all pin discrepancies
are below1e-12, a uniform1e-8 additive margin in each master bound
dominates this finite-support adjustment and its force/Hessian
changes. The working radius.009 leaves room within the certified
.01 tubes. This correction must be stated; merely treating the
floating pin values as equal would not define the claimed family.
\end{Verbatim}

\bibliographystyle{unsrt}
\bibliography{bibliography/references}

@article{thomson1971,
  author = {R. Thomson and C. Hsieh and V. Rana},
  title = {Lattice trapping of fracture cracks},
  journal = {Journal of Applied Physics},
  year = {1971},
  volume = {42},
  pages = {3154--3160},
  doi = {10.1063/1.1660699},
  note = {\href{https://doi.org/10.1063/1.1660699}{doi:10.1063/1.1660699}},
}

@article{buze2020,
  author = {M. Buze and T. Hudson and C. Ortner},
  title = {Analysis of cell size effects in atomistic crack propagation},
  journal = {ESAIM: Mathematical Modelling and Numerical Analysis},
  year = {2020},
  volume = {54},
  pages = {1821--1847},
  doi = {10.1051/m2an/2020005},
  note = {\href{https://doi.org/10.1051/m2an/2020005}{doi:10.1051/m2an/2020005}},
}

@article{buze2021,
  author = {M. Buze and J. R. Kermode},
  title = {Numerical-continuation-enhanced flexible boundary condition scheme applied to mode-{I} and mode-{III} fracture},
  journal = {Physical Review E},
  year = {2021},
  volume = {103},
  pages = {033002},
  doi = {10.1103/PhysRevE.103.033002},
  note = {\href{https://doi.org/10.1103/PhysRevE.103.033002}{doi:10.1103/PhysRevE.103.033002}},
}

@article{braun2025,
  author = {J. Braun and M. Buze},
  title = {Incompleteness of {Sinclair}-type continuum flexible boundary conditions for atomistic fracture simulations},
  journal = {Multiscale Modeling \& Simulation},
  year = {2025},
  volume = {23},
  pages = {711--752},
  doi = {10.1137/24M1661078},
  note = {\href{https://doi.org/10.1137/24M1661078}{doi:10.1137/24M1661078}},
}
\end{document}